\RequirePackage{fix-cm}

\documentclass[smallextended]{svjour3}       
\usepackage{etoolbox}
 \newcommand{\makeheadbox}{}
\smartqed  
\usepackage[colorlinks=true,linkcolor=blue,citecolor=blue,urlcolor=blue]{hyperref}
\usepackage{siunitx}
\usepackage{graphicx}
\usepackage{lipsum}
\usepackage{multirow}%
\usepackage{amsmath,amssymb,amsfonts}%
\usepackage{mathrsfs}%
\usepackage[title]{appendix}%
\usepackage{xcolor}%
\usepackage{textcomp}%
\usepackage{manyfoot}%
\usepackage{booktabs}%
\usepackage{algorithm}%
\usepackage{algorithmicx}%
\usepackage{algpseudocode}%
\usepackage{listings}%
\usepackage[utf8]{inputenc} 
\usepackage[T1]{fontenc}
\usepackage{lmodern} 
\usepackage[cal=rsfs]{mathalfa}
\usepackage{url}
\usepackage{ifthen}
\usepackage{graphicx}
\usepackage{amssymb}
\usepackage{mathrsfs,bm,braket,marginnote}
\usepackage{enumerate}
\usepackage{lipsum}
\usepackage{alltt}
\usepackage{dcolumn}
\usepackage{multirow}
\usepackage{stackrel}
\usepackage{longtable}
\usepackage{tikz}
\usepackage{cite}
\usetikzlibrary{arrows,calc}
\usepackage{verbatim}
\usepackage{breqn}
\usepackage{chngcntr} 
\usetikzlibrary{decorations.pathreplacing,decorations.pathmorphing}
\usepackage{mathtools} 
\usepackage{etoolbox}

\reversemarginpar
\newcommand{\e}{{\rm e}}
\newcommand{\tr}{{\mathrm{tr}}}

\newcommand{\dd}{\,\mathrm{d}}%
\newtheorem{prop}{Proposition}
\newlength{\myeqnumwidth}
\newcommand{\bignumber}{%
\llap{\makebox[\myeqnumwidth][l]{(\refstepcounter{equation}\theequation)\quad}}%
}

\usepackage{refcount}

\begin{document}
\title{Skewness of von Neumann Entropy over Bures-Hall Random States}


\author{Linfeng Wei \and Youyi Huang \and Lu Wei}

\institute{Linfeng Wei \email{linwei@ttu.edu}  
\and Lu Wei \email{luwei@ttu.edu} \at Department of Computer Science, Texas Tech University, Lubbock, TX 79409, USA
\and Youyi Huang \email{yhuang@ucmo.edu} \at Department of Computer Science and Cybersecurity, University of Central Missouri, Warrensburg, MO 64093, USA \\}
\date{\vspace{-14ex}}
\maketitle

\begin{abstract}
We study the degree of entanglement, as measured by von Neumann entropy, of bipartite systems over the Bures-Hall ensemble. Closed-form expressions of the first two cumulants of von Neumann entropy over the ensemble have been recently derived in the literature. In this paper, we focus on its skewness by calculating the third cumulant that describes the degree of asymmetry of the distribution. The main result is an exact closed-form formula of the third cumulant, which leads to a more accurate approximation to the distribution of von Neumann entropy. The key to obtaining the result lies on finding a dozen of new summation identities in simplifying a large number of finite summations involving polygamma functions.

\end{abstract}

\newpage

\section{Introduction}
We study the statistical behavior of entanglement in quantum bipartite systems over Bures-Hall ensemble~\cite{Hall98,Zyczkowski01,Sommers03,Sommers04,Osipov10,Borot12,FK16,Bertola14,Sarkar19,Wei20BHA,Wei20BH,LW21}, which is one of the three main generic state ensembles, together with Hilbert-Schmidt ensemble~\cite{Page93,Foong94,Ruiz95,HLW06,VPO16,Wei17,Wei20,HWC21,Lubkin78,Sommers04,Giraud07,MML02,Wei22,huang2025cumulantstructuresentanglemententropy} and fermionic Gaussian ensemble~\cite{OGMM19,LRV20,LRV21,BHK21,ST21,BHKRV22,HW22,huang_entropy_2023,HWISIT}. In particular, the degree of entanglement over such ensembles can be measured by the von Neumann entropy, which is a linear statistic described by its cumulants/moments. In the literature, the mean and variance of von Neumann entropy over Bures-Hall ensemble have been obtained~\cite{Sarkar19,Wei20BHA,Wei20BH}. In this work, we continue the investigation of its skewness by deriving an exact closed-form formula of the third cumulant. 

The bipartite system of the Bures-Hall ensemble is described as follows. Consider two Hilbert spaces $\mathcal{H}_{A}$ of dimension $m$ and $\mathcal{H}_{B}$ of dimension $n$, the Hilbert space $\mathcal{H}_{A+B}$ of the composite system is given by the tensor product of the subsystems $\mathcal{H}_{A}$ and $\mathcal{H}_{B}$, i.e., $\mathcal{H}_{A+B}=\mathcal{H}_{A}\otimes\mathcal{H}_{B}$. A random pure state of the composite system $\mathcal{H}_{A+B}$ is defined as a linear combination of the random coefficients $z_{i,j}$ and the complete basis $\big\{\Ket{i^{A}}\big\}$ and $\big\{\Ket{j^{B}}\big\}$ of $\mathcal{H}_{A}$ and $\mathcal{H}_{B}$, 
\begin{equation}
\Ket{\psi_0}=\sum_{i=1}^{m}\sum_{j=1}^{n}z_{i,j}\Ket{i^{A}}\otimes\Ket{j^{B}},  
\end{equation}
where each $z_{i,j}$ follows the standard complex Gaussian distribution of zero mean and unit variance with the probability constraint $\sum_{i,j}|z_{i,j}|^2=1$. We then consider a superposition of the above state as~\cite{Sarkar19}
\begin{equation}\label{eq:SB}
\Ket{\varphi}=\Ket{\psi_0}+\left(\mathbf{U}\otimes\mathbf{I}_{m}\right)\Ket{\psi_0},
\end{equation}
where $\mathbf{U}$ is an $m\times m$ unitary random matrix with the measure proportional to $\det^{2\alpha+1}\left(\mathbf{I}_{m}+\mathbf{U}\right)$, where the parameter $\alpha$ takes half-integer values as
\begin{equation}\label{eq:aBH}
\alpha=n-m-\frac{1}{2}.
\end{equation} The corresponding density matrix of the pure state~(\ref{eq:SB}) is
$\rho=\Ket{\varphi}\Bra{\varphi}$,
which has the natural probability constraint $\tr(\rho)=1$.
We assume without loss of generality that $m\leq n$, i.e., the dimension of subsystem $A$ is less than or equal to dimension of subsystem $B$. The reduced density matrix $\rho_{A}$ of subsystem $A$ is computed by partial tracing of the full density matrix over the other subsystem $B$ as $\rho_{A}=\tr_{B}(\rho)$.
The resulting density of eigenvalues of $\rho_{A}$, $\lambda_{i}\in[0,1]$, $i=1,\dots,m$, is the Bures-Hall measure~\cite{Hall98,Zyczkowski01,Sommers03,Sarkar19}
\begin{equation}\label{eq:BH}
f\left(\bm{\lambda}\right)=\frac{1}{C}~\delta\left(1-\sum_{i=1}^{m}\lambda_{i}\right)\prod_{1\leq i<j\leq m}\frac{\left(\lambda_{i}-\lambda_{j}\right)^{2}}{\lambda_{i}+\lambda_{j}}\prod_{i=1}^{m}\lambda_{i}^{\alpha},
\end{equation}
where the constant $C$ is
\begin{equation}\label{eq:cBH}
C=\frac{2^{-m(m+2\alpha)}\pi^{m/2}}{\Gamma\left(m(m+2\alpha+1)/2\right)}\prod_{i=1}^{m}\frac{\Gamma(i+1)\Gamma(i+2\alpha+1)}{\Gamma(i+\alpha+1/2)}
\end{equation}
with $\delta(x)$ denoting the Dirac delta function and $\Gamma(x)$ being the Gamma function. The von Neumann entropy of the bipartite system is defined as 
\begin{equation}\label{eq:vN}
S=-\tr\left(\rho_{A}\ln\rho_{A}\right)=-\sum_{i=1}^{m}\lambda_{i}\ln\lambda_{i},
\end{equation}
supported in $S\in\left[0,\ln{m}\right]$, which degenerates to the separable state $S=0$ when 
\begin{equation}
\lambda_{1}=1,~~\lambda_{2}=\dots=\lambda_{m}=0
\end{equation}
and achieves the maximally-entangled state $S=\ln{m}$ when 
\begin{equation}
\lambda_{1}=\dots=\lambda_{m}=\frac{1}{m}.
\end{equation}
Statistical information of the von Neumann entropy as an entanglement measure is revealed by its cumulants: the first cumulant (mean value) implies the typical degree of entanglement, the second cumulant (variance) specifies the fluctuation around the typical value, and the higher order cumulants such as skewness describe the tail distribution. 

While the calculation of entropy cumulants relies heavily on the summation form of correlation functions, the final results are always in compact closed form without summations~\cite{HW22,huang_entropy_2023,HWISIT,HWC21,huang2025cumulantstructuresentanglemententropy,Wei17,Wei19,Wei20,Wei20BHA,Wei20BH}. This is the case, for instance, of the mean~\cite{Wei20BHA} and the variance~\cite{Wei20BH} of von Neumann entropy over the Bures-Hall ensemble. However, various summations are not simplifiable to closed-form expressions themselves that occur in the procedure, and the simplification has always been a challenging and case-by-case task. Specifically, as the order of cumulant increases, the summation forms of cumulant integrals become more complicated, where the $k$-th cumulant involves up to multiple summations of $k$ nested indices. At the same time, the number of unsimplifiable single-sums involving polygamma functions grows rapidly as cumulant order increases. We refer to such single-sums as anomalies~\cite{huang2025cumulantstructuresentanglemententropy} since they will be canceled in the end, but appear in the intermediate step of the simplification. 

In the current computation of the third cumulant, we encountered new types of anomalies shown in Table~\ref{tab:basis} compared to the anomalies in variance computation~\cite{Wei20BH} provided in Table~\ref{tab:exist}. The new anomalies often come with more intricate forms with increasing number of arguments. The key to obtaining the main result in Proposition~\ref{p:1} lies in the systematic cancellation of the additional twelve new  anomalies, along with the derivation of underlying new summation identities. Finally, we note that whilst a summation-free framework for cumulant calculation is being established for the Hilbert-Schmidt ensemble~\cite{huang2025cumulantstructuresentanglemententropy}, in this work, we follow the summation-and-cancellation framework for the Bures-Hall ensemble as in~\cite{Wei20BHA,Wei20BH}.

The rest of the paper is organized as follows. In Section~\ref{sec2}, we present the main result of a closed-form third cumulant expression of von Neumann entropy before discussing the new challenge in simplifying summations. In Section \ref{sec:T3}, we provide detailed calculations of the main result, where summation identities for the simplification are listed in Appendix~\ref{secA1}.

\section{Main Results and New Challenges in Simplification}\label{sec2}
\subsection{Exact third cumulant formula and numerical results}\label{sec2.1}
To present the cumulant results, we first set up the following notations. Denote the $k$-th order polygamma function as~\cite{Prudnikov86}
\begin{equation}\label{eq:polygamma}
\psi_{k}(z)=\frac{\dd^{k+1}}{\dd z^{k+1}}\ln\Gamma(z)=\frac{\dd^{k}}{\dd z^{k}}\psi_{0}(z)
\end{equation}
that satisfies 
\begin{equation}\label{eq:psisre}
\psi_{k}(z+n)-\psi_{k}(z)=(-1)^{k}k!\sum_{i=0}^{n-1}\frac{1}{(z+i)^{k+1}}.
\end{equation}
Polygamma functions with positive integer argument $l$ admit finite-sum representations
\begin{subequations}
\begin{eqnarray}
\psi_0(l)&=&-\gamma+\sum_{\smash{i=1}}^{l-1}\frac{1}{i}\\
\psi_k(l)&=&(-1)^{k+1} k!\left(\zeta(k+1)-\sum_{\smash{i=1}}^{l-1}\frac{1}{i^{k+1}}\right),~~~k\geq 1,
\end{eqnarray}
\end{subequations}
with $\gamma \approx 0.5772$ being the Euler's constant and $\zeta(s)=\sum_{\smash{i=1}}^{\infty}\frac{1}{i^s}$ being the Riemann zeta function. We also need finite-sum representations of polygamma functions with half-integer arguments
\begin{subequations}
\begin{eqnarray}
\!\!\!\!\!\!\psi_0\!\left(l+\frac{1}{2}\right)&=&-\gamma-2\ln2+2\sum_{i=0}^{l-1}\frac{1}{2i+1} \\
\!\!\!\!\!\!\psi_{k}\!\left(l+\frac{1}{2}\right)&=&(-1)^{k+1}k!\!\left(\frac{\zeta(k+1)}{\left(2^{k+1}-1\right)^{-1}}-\sum_{i=0}^{l-1}\frac{2^{k+1}}{(2i+1)^{k+1}}\right),~k\geq 1.
\end{eqnarray}
\end{subequations}

With the notations above, we have the first three cumulants formulas as follows. The exact mean~\cite{Sarkar19, Wei20BHA} and variance~\cite{Wei20BH} of von Neumann entropy~(\ref{eq:vN}) over the Bures-Hall ensemble~(\ref{eq:BH}) are
\begin{equation}\label{eq:vNm}
\kappa_1=\psi_{0}\!\left(mn-\frac{m^2}{2}+1\right)-\psi_{0}\!\left(n+\frac{1}{2}\right)
\end{equation}
and
\begin{eqnarray}\label{eq:vNv}
\kappa_2&=&-\psi_{1}\!\left(mn-\frac{m^2}{2}+1\right) +\frac{2n(2n+m)-m^{2}+1}{2n(2mn-m^2+2)}\psi_{1}\!\left(n+\frac{1}{2}\right),
\end{eqnarray}
respectively. 

The main result of this work is presented in Proposition~\ref{p:1} below.
\begin{prop}\label{p:1} 
The exact third order cumulant of von Neumann entropy~(\ref{eq:vN}) over the Bures-Hall ensemble~(\ref{eq:BH}) is
\begin{equation}\label{eq:main}
\kappa_3=\psi_2\!\left(mn-\frac{m^2}{2}+1\right)+a_{1}\psi_2\!\left(n+\frac{1}{2}\right)+a_2\psi_1\!\left(n+\frac{1}{2}\right),
\end{equation}
where
\begin{eqnarray}
a_1 &=&\frac{4m^2-8mn-4n^2-7}{\left(2mn-m^2+2\right)\left(2mn-m^2+4\right)} \\
a_2 &=&\frac{2(m^2-1)\left((m-2n)^2-1\right)\left(-2m^2+4 m n-12 n^2+7\right)}{n\left(2 m n-m^2+2\right)^2\left(2 m n-m^2+4\right)\left(4 n^2-1\right)}.
\end{eqnarray}
\end{prop}
The proof of Proposition~\ref{p:1} is in Section~\ref{sec:T3}.\\

The above third order cumulant formula leads to the skewness $\gamma_1$ as
\begin{equation}\label{skewness}
\gamma_1=\mathbb{E}\left[\left(\frac{S-\kappa_1}{\kappa_2}\right)^3\right]=\frac{\kappa_3}{\kappa_2^{3/2}},
\end{equation}
which provides a better understanding of the distribution of von Neumann entropy. To illustrate this, we consider standardized von Neumann entropy
\begin{equation}\label{eq:X}
X=\frac{S-\kappa_1}{\sqrt{\kappa_2}}
\end{equation}
so that the mean and variance of the random variable $X$ become
\begin{equation}
\kappa_1^{(X)}=0,~~~~\kappa_2^{(X)}=1.
\end{equation}
The obtained formula $\kappa_3$ is incorporated as a correction term~\cite{Wei20} in approximating the distribution of standardized von Neumann entropy $X$ as
\begin{equation}\label{eq:fT3}
f_X(x)\approx\phi(x)\left(1+\frac{\kappa_3}{6\kappa_2^{3/2}}\left(x^3-3x\right)\right),
\end{equation}
where
\begin{equation}\label{eq:iappr}
\phi(x)=\frac{1}{\sqrt{2\pi}}\mathrm{e}^{-\frac{1}{2}x^2}, \qquad x\in(-\infty,\infty),
\end{equation} 
denotes the standard Gaussian density.
\begin{figure}[ht]
\centering
\includegraphics[width=0.85\textwidth]{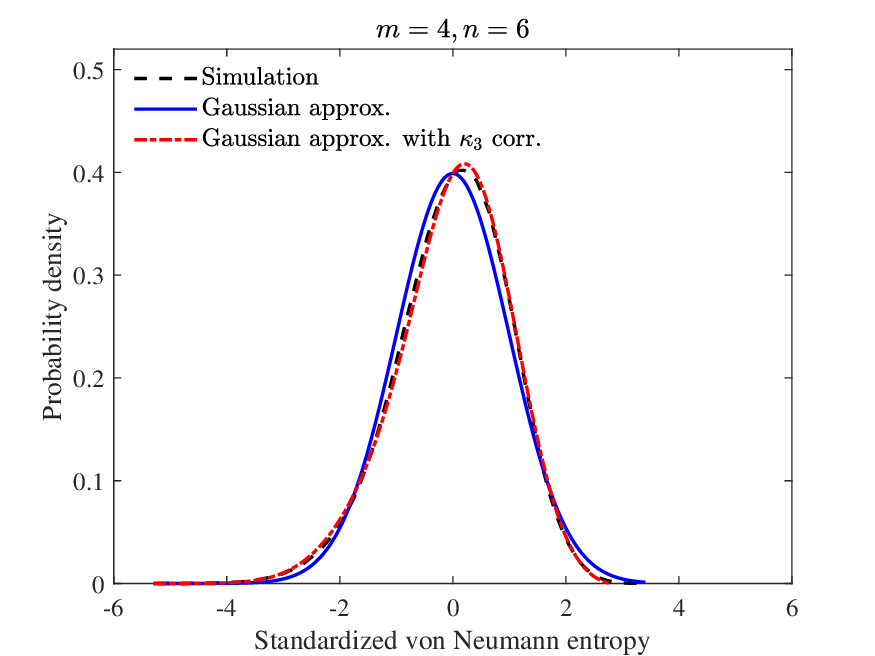}
\caption{Probability densities of standardized von Neumann entropy: A comparison of Gaussian density~(\ref{eq:iappr}) with and without skewness correction to simulation results. The dashed curve in black refers to the simulation of standardized von Neumann entropy~(\ref{eq:X}) of subsystem dimensions $m=4$, $n=6$. The solid blue curve represents the Gaussian density~(\ref{eq:iappr}). The dash-doted curve in red refers to the Gaussian density~(\ref{eq:iappr}) with correction of our third cumulant result~Proposition \ref{p:1}.}\label{fig1}
\end{figure}
\begin{figure}[ht]
\centering
\includegraphics[width=0.85\textwidth]{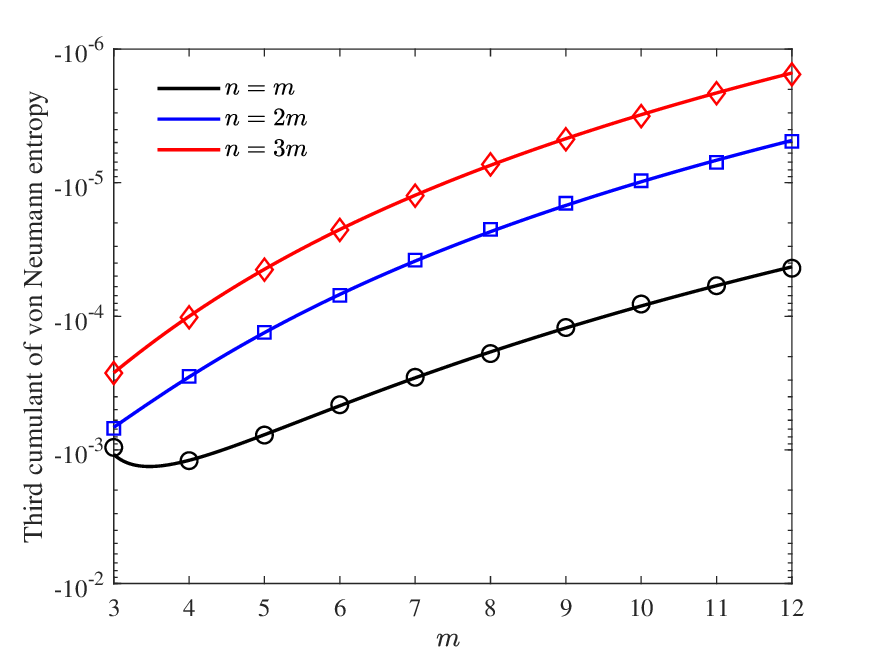}
\caption{Third order cumulant of von Neumann entropy: Log-Linear plot of the result~(\ref{eq:main}) from Proposition~\ref{p:1} and numerical simulations for integer $m$ between $3$ to $12$ when $n=m$, $n=2m$, and $n=3m$. The black, blue, and red curves are results from Proposition \ref{p:1} when $n=m$, $n=2m$, and $n=3m$, respectively, whereas the black circle, blue square, and red diamond are the points on simulation for integer $m$ between $3$ to $12$ when $n=m$, $n=2m$, and $n=3m$, respectively.}\label{fig2}
\end{figure}As shown in Figure~\ref{fig1}, by incorporating the additional knowledge of skewness, the resulting approximation~(\ref{eq:fT3}) becomes more accurate than the Gaussian approximation~(\ref{eq:iappr}). As a sanity check, we also compare the $\kappa_3$ formula~(\ref{eq:main}) with numerical simulations as shown in Figure~\ref{fig2}.

It is worthwhile mentioning the conjecture in~\cite{Wei20BH} that, in the limit
\begin{equation}\label{eq:lim}
m\to\infty,~~~~n\to\infty,~~~~\frac{m}{n}=c\in(0,1],
\end{equation}
the standardized von Neumann entropy~(\ref{eq:X}) converges in distribution to a Gaussian random variable with zero mean and unit variance. The obtained closed-form result in Proposition \ref{p:1} allows us to analyze the third cumulant of standardized von Neumann entropy through asymptotic behavior of polygamma functions
\begin{equation}
\psi_j(x) = \Theta\left(\frac{1}{x^j}\right),~~~ x\rightarrow \infty, ~~~j\geq1, 
\end{equation}
where $\Theta(\cdot)$ is the big-theta notation of Bachmann-Landau symbols. Taking $m=cn$ in the results~(\ref{eq:vNv}) and~(\ref{eq:main}), we have, in the limiting regime~(\ref{eq:lim}),
\begin{eqnarray}
\kappa_2&=&\Theta\left(\frac{1}{n^2}\right)\\
\kappa_3&=&\Theta\left(\frac{1}{n^4}\right),
\end{eqnarray}
which leads to 
\begin{equation}
\kappa_3^{(X)}=\frac{\kappa_3}{\kappa_2^{3/2}}=\frac{\Theta\left(\frac{1}{n^4}\right)}{\Theta\left(\frac{1}{n^3}\right)}=\Theta\left(\frac{1}{n}\right)
\end{equation}
under the same limit~(\ref{eq:lim}). This shows that the third cumulant of standardized von Neumann entropy $\kappa_3^{(X)}$ indeed vanishes in the regime~(\ref{eq:lim}), providing a further evidence to the aforementioned conjecture.
      
\subsection{New summation identities and cancellation framework}\label{sec2.2}
The main effort in obtaining the closed-form expression~(\ref{eq:main}) lies in simplifying summations arising from the integrals in Section~\ref{sec3.2}, where many of the single sums involving rational functions of polygamma functions occurred in each piece of summation themselves may not be further simplified into closed forms. We refer to such unsimplifiable summations as anomalies~\cite{huang2025cumulantstructuresentanglemententropy} since after collecting these terms together, they cancel completely. Some of the anomalies appeared in existing works, while others are new, which are respectively listed in Table~\ref{tab:exist} and Table~\ref{tab:basis} below, where $a$ takes the values $m$, $\alpha+m$, $2\alpha+m$, $2\alpha+2m$; $b,c $ takes the values $ 0$, $\alpha$, $2\alpha$, $2\alpha+m$, and $b\neq c$ in $\Omega_6^{(b,c)}$, whereas the $b=c$ case leads to a closed form.
 
\begin{table}[htbp]
\renewcommand{\arraystretch}{3}
\caption{Anomalies in existing works~\cite{HW22,huang_entropy_2023,HWISIT,HWC21,huang2025cumulantstructuresentanglemententropy,Wei17,Wei19,Wei20,Wei20BHA,Wei20BH}} \label{tab:exist}
\begin{tabular}{l l}
\hline
$\displaystyle{\Omega_1^{(a)}\!=\!   \sum_{\smash{k=1}}^{m} \frac{\psi_0(k)}{a+1-k}}$&$    \displaystyle{  \Omega_2^{(a)}\!=\!  \sum_{\smash{k=1}}^{m} \frac{\psi_0(a+1-k)}{k}}$ \\
$\displaystyle{\Omega_3^{(b,c)}\!=\!  \sum_{\smash{k=1}}^{m} \frac{\psi_0(k+b)}{(k+c)^2}}$  &$\displaystyle{   \Omega_4^{(b,c)}\!=\!\sum_{\smash{k=1}}^{m} \frac{\psi_0^2(k+b)}{k+c}}$\\
$\displaystyle{\Omega_5^{(b,c)}\!=\! \sum_{\smash{k=1}}^{m} \frac{\psi_1(k+b)}{k+c}}$&$\displaystyle{   \Omega_6^{(b,c)}\!=\!\sum_{\smash{k=1}}^{m} \frac{\psi_0(k+b)}{k+c} }$ \vspace{0.1cm}\\
\hline
\end{tabular}
\end{table}
\begin{table}[htbp]
\renewcommand{\arraystretch}{3}
\caption{New anomalies in the current work}\label{tab:basis}
\begin{tabular}{l l}
\hline
$ \displaystyle{  \Omega_7^{(a)}\!=\!  \sum_{\smash{k=1}}^{m} \frac{\psi_0(k)}{(a+1-k)^2}}$&$\displaystyle{   \Omega_8^{(a)}\!=\!\sum_{\smash{k=1}}^{m} \frac{\psi_0^2(k)}{a+1-k}}$\\
$\displaystyle{\Omega_{9}^{(a)}\!=\!\sum _{\smash{k=1}}^m \frac{\psi_0 (a+1-k)}{k^2}}$&$ \displaystyle{   \Omega_{10}^{(a)}\!=\! \sum _{\smash{k=1}}^m \frac{\psi_0^2 (a+1-k)}{k}}$  \\
 $  \displaystyle{    \Omega_{11}^{(a)}\!=\!\sum_{\smash{k=1}}^{m} \frac{\psi_0(k)\psi_0(k+a-m)}{m+1-k}}$&$ \displaystyle{     \Omega_{12}^{(a)}\!=\! \sum_{\smash{k=1}}^{m} \frac{\psi_0(k)\psi_0(a+1-k)}{m+1-k}}$ \\
$    \displaystyle{  \Omega_{13}^{(a)}\!=\! \sum_{\smash{k=1}}^{m} \frac{\psi_0(k)\psi_0(a+1-k)}{a+1-k} }$&$\displaystyle{   \Omega_{14}^{(a)}\!=\! \sum_{\smash{k=1}}^{m} \frac{\psi_0(k)\psi_0(a+1-k)}{k}} $ \\
$\displaystyle{   \Omega_{15}^{(b)}\!=\! \sum_{\smash{k=1}}^{m} \frac{\psi_0(k)\psi_0(k+b)}{k+b}}$& $\displaystyle{  \Omega_{16}^{(b)}\!=\! \sum_{\smash{k=1}}^{m} \frac{\psi_0(k)\psi_0(k+b)}{k}}$\\
          $\displaystyle{   \Omega_{17}^{(a)}\!=\!  \sum_{\smash{k=1}}^{m} \frac{\psi_1(k)}{a+1-k}}$&$\displaystyle{ \Omega_{18}^{(a)}\!=\!  \sum_{\smash{k=1}}^{m} \frac{\psi_1(a+1-k)}{k}}$ \vspace{0.1cm}\\
\hline
\end{tabular}
\end{table}

Similar phenomenon of the arising and eventual cancellation of unsimplifiable single sum anomalies also appears in existing works~\cite{HW22,huang_entropy_2023,HWISIT,HWC21,huang2025cumulantstructuresentanglemententropy,Wei17,Wei19,Wei20,Wei20BHA,Wei20BH}, where most of such anomalies show up in pairs with coefficients opposite to each other so that are added up in summation to cancel immediately. However, when it comes to third order cumulant that consists of triple sums, the individual single sum anomalies are different compared to existing ones in two ways. Firstly, the summand of single sums derived from triple sums may have two different kinds of polygamma functions multiplying together on the numerator of the rational function, one being positive and one being negative in summation variable, e.g., $\Omega_{12}^{(a)}$ to $\Omega_{14}^{(a)}$. As a comparison, the anomalies with negative summation variables so far in existing works~\cite{HW22,huang_entropy_2023,HWISIT,HWC21,huang2025cumulantstructuresentanglemententropy,Wei17,Wei19,Wei20,Wei20BHA,Wei20BH} consist of only one such kind of polygamma function on the numerator, e.g., 
\begin{equation}
\sum\limits_{\smash{k=1}}^{m}\frac{\psi_0 (a+1-k)}{k}.
\end{equation} 
Secondly, the parameter $a$ in the argument of polygamma functions observes more values, e.g., $\Omega_{7}^{(a)}$ to $\Omega_{14}^{(a)}$, $\Omega_{17}^{(a)}$, $\Omega_{18}^{(a)}$ for $a\neq m$, where the $a=m$ cases of $\Omega_{7}^{(a)}$ to $\Omega_{11}^{(a)}$ fall back to existing single-sum anomalies by re-ordering the summations from $k$ to $m+1-k$, e.g.,
\begin{equation}
\Omega_{11}^{(m)}=\sum\limits_{\smash{k=1}}^{m}\frac{\psi_0^2(m+1-k)}{k},
\end{equation} or a pair of the same kind anomaly, e.g., 
\begin{eqnarray}
\Omega_{7}^{(m)}&=&\Omega_{9}^{(m)}=\sum\limits_{\smash{k=1}}^m \frac{\psi_0 (m+1-k)}{k^2} \\
\Omega_{8}^{(m)}&=&\Omega_{10}^{(m)}=\sum\limits_{\smash{k=1}}^m \frac{\psi_0 ^2(m+1-k)}{k}.
\end{eqnarray} 
Therefore, the single sum anomalies arise in Table~\ref{tab:basis} serve as an extended version of rational function involving polygamma functions from the triple sums of the cumulant integrals, and the existing single sum anomalies from literature perform as degenerated special cases. As a consequence, unlike the cancellation by finding pairs of the same anomaly with opposite coefficients in existing works, the possible cancellation of anomalies now relies on them being transferred to the same kind. Hence, a full and complete investigation of the relations of such anomalies is required in order to perform cancellation systematically before arriving at the closed-form expression~(\ref{eq:main}). We refer to Appendix~\ref{secA1} for identities to relate such single-sum anomalies, where all summation anomalies are transformed into basic ones $\Omega_{4}^{(b,c)}$, $\Omega_{5}^{(b,c)}$, $\Omega_{6}^{(b,c)}$, and $\Omega_{10}^{(a)}$, plus some closed-form terms. We also provide examples of anomalies adding up to closed forms in Appendix~\ref{secA1}.

\section{Calculation of Third Cumulant}\label{sec:T3}
\subsection{Moment conversion and correlation functions}\label{sec:3.1}
To calculate the cumulants/moments over the Bures-Hall ensemble~(\ref{eq:BH}), a standard way is to calculate these over the unconstrained Bures-Hall ensemble \cite{Sarkar19,Wei20BHA,Osipov10,Wei20BH,LW21}, whose density function reads 
\begin{equation}\label{eq:BHu}
h\left(\bm{x}\right)=\frac{1}{C'}\prod_{1\leq i<j\leq m}\frac{\left(x_{i}-x_{j}\right)^{2}}{x_{i}+x_{j}}\prod_{i=1}^{m}x_{i}^{\alpha}\e^{-x_{i}},
\end{equation}
where $x_{i}\in[0,\infty)$, $i=1,\dots,m$, and the constant $C'$ is $C'=C~\Gamma\left(d\right)$ with 
\begin{equation}
d=\frac{1}{2}m\left(m+2\alpha+1\right). 
\end{equation}
Such an unconstrained ensemble is related to the original Bures-Hall ensemble~(\ref{eq:BH}) by the factorization~\cite{Wei20BHA,Wei20BH}
\begin{equation}\label{eq:g2ft}
h(\bm{x})\prod_{i=1}^{m}\dd x_{i}=f(\bm{\lambda})g(\theta)\dd\theta\prod_{i=1}^{m}\dd\lambda_{i},
\end{equation}
where 
\begin{equation}
g(\theta)=\frac{1}{\Gamma\left(d\right)}\e^{-\theta}\theta^{d-1}
\end{equation}is the density of trace of the unconstrained ensemble 
\begin{equation}\label{eq:tr}
\theta=\sum_{i=1}^{m}x_{i}~~~~\theta\in[0,\infty).
\end{equation}
The factorization~(\ref{eq:g2ft}) implies that the random variable $\theta$ is independent of $\bm{\lambda}$, and therefore independent of $S$. The independency allows us to perform the change of variable 
\begin{equation}\label{eq:S3T}
S^3=-\theta^{-3} T^3+3S^2\ln \theta -3S \ln^{2} \theta+ \ln^{3} \theta,
\end{equation}
where 
\begin{equation}\label{eq:T}
T=\sum_{i=1}^{m}x_{i}\ln x_i
\end{equation}
is the von Neumann entropy of the unconstrained Bures-Hall ensemble~(\ref{eq:BHu}). By the fact that
\begin{equation}
1=\int_{0}^{\infty}\frac{1}{\Gamma\left(d+3\right)}\e^{-\theta}\theta^{d+2}\dd\theta,
\end{equation}
we deduce the moment conversion
\begin{eqnarray}\label{mc}
\mathbb{E}_f\!\left[S^3\right]&=&\int_{0}^{\infty}\int_{\bm{\lambda}}\frac{\e^{-\theta}\theta^{d+2}}{\Gamma\left(d+3\right)}S^{3}f(\bm{\lambda})\dd\theta\prod_{i=1}^{m}\dd\lambda_{i} \nonumber \\
&=&-\frac{1}{(d)_3}\mathbb{E}_h\!\left[T^3\right]+3\psi_0(d+3)\mathbb{E}_f\!\left[S^2\right]-3\left(\psi_1(d+3)+\psi_0^2(d+3)\right) \nonumber \\
&&\times\mathbb{E}_f\!\left[S\right]+\left(\psi_2(d+3)+3\psi_1(d+3)\psi_0(d+3)+\psi_0^3(d+3)\right),
\end{eqnarray}
where $(a)_n=\Gamma(a+n)/\Gamma(a)$ denotes the Pochhammer's symbol and we have utilized~(\ref{eq:g2ft}) and~(\ref{eq:S3T}), along with the derivatives of the Gamma function 
\begin{equation}
\Gamma(a)=\int_{0}^{\infty}\!\!\e^{-\theta}\theta^{a-1}\dd\theta.
\end{equation}

We label the first three cumulants of the von Neumann entropy of unconstrained Bures-Hall ensemble as $\kappa_1^T, \kappa_2^T$, and $\kappa_3^T$. Along with results of $\kappa_1^T$, $\kappa_2^T$ obtained in~\cite{Sarkar19, Wei20BHA,Wei20BH} and~(\ref{eq:vNm}), (\ref{eq:vNv}), the calculation of $\kappa_3$ now boils down to calculation of $\kappa_3^T$. The $k$-point ($k\leq m$) correlation function $\rho_{k}(x_{1},\dots,x_{k})$ of the unconstrained ensemble is known to follow a Pfaffian point process of a $2k\times2k$ antisymmetric matrix~\cite{FK16}, where the corresponding correlation kernels are related to those of the Cauchy-Laguerre biorthogonal ensemble~\cite{Bertola14}. As a result, the computation of moments/cumulants of linear statistics of the Bures-Hall ensemble can be performed over the kernels of the Cauchy-Laguerre ensemble, where the four correlation kernels~\cite{FK16, LW21} are $K_{00}(x,y)$, $K_{01}(x,y)$, $K_{10}(x,y)$, and $K_{11}(x,y)$. In particular, the first three moments of von Neumann entropy of unconstrained Bures-Hall ensemble~(\ref{eq:BHu}) can be written in terms of first three point densities as
\begin{eqnarray}
\mathbb{E}_{h}\!\left[T\right]&=& m\int_{0}^{\infty}f(x)~h_{1}(x)\dd x\label{eq:ET1} \\
\mathbb{E}_{h}\!\left[T^{2}\right]&=& m\int_{0}^{\infty}\!\!f^2(x)h_{1}(x)\mathrm{d} x+(m-1)_2\!\int_{0}^{\infty}\!\!\int_{0}^{\infty}\!\!f^2(x)f(y)h_{2}\left(x,y\right)\mathrm{d} x \mathrm{d} y \\\label{eq:ET2} 
\mathbb{E}_{h}\!\left[T^{3}\right]&=& m\int_{0}^{\infty} f^3(x)~h_{1}(x)\mathrm{d} x+3(m-1)_2\int_{0}^{\infty}\int_{0}^{\infty}f^2(x)f(y)h_{2}(x,y)\mathrm{d} x \mathrm{d} y\nonumber\\
&&+(m-2)_3\int_{0}^{\infty}\int_{0}^{\infty}\int_{0}^{\infty}f(x)f(y)f(z)h_{3}(x,y,z)\mathrm{d} x \mathrm{d} y\mathrm{d} z, \label{eq:ET3}
\end{eqnarray} 
where we denote
\begin{equation}
f(x) = x\ln x. 
\end{equation}
Here, the density functions $h_1(x), h_2(x,y)$, and $h_3(x,y,z)$ are~\cite{LW21,FK16}
\begin{eqnarray}
h_{1}(x)&=&\frac{1}{2m}\left(K_{01}(x,x)+K_{10}(x,x)\right)\label{eq:h1} \\
h_{2}(x,y)&=&\frac{1}{4m(m-1)}(\left(K_{01}(x,x)+K_{10}(x,x)\right)\left(K_{01}(y,y)+K_{10}(y,y)\right)\nonumber \\
&&-2K_{01}(x,y)K_{01}(y,x) -2K_{10}(x,y)K_{10}(y,x)\nonumber \\
&&-2K_{00}(x,y)K_{11}(x,y)-2K_{00}(y,x)K_{11}(y,x))\label{eq:h2} \\
h_{3}(x,y,z)&=&\frac{1}{8m(m-1)(m-2)}\left(h_{\text{A}}+h_{\text{B}}+h_{\text{C}}+h_{\text{D}}\right), \label{eq:h3}
\end{eqnarray}
where explicit expressions of $h_{\text{A}}$, $h_{\text{B}}$, $h_{\text{C}}$, $h_{\text{D}}$ are listed in Appendix~\ref{HAHD}. Together with the standard moment-cumulant relation of a random variable $X$,
\begin{subequations}\label{eq:m2c}
\begin{align}
\kappa_1(X)&=\mathbb{E}[X] \\
\kappa_2(X)&=\mathbb{E}\left[X^2\right]-\mathbb{E}^2[X] \\
\kappa_3(X)&=\mathbb{E}\left[X^3\right]-3\mathbb{E}\left[X^2\right]\mathbb{E}[X]+2\mathbb{E}^3[X],
\end{align}
\end{subequations}
the third cumulant of von Neumann entropy of the unconstrained ensemble~(\ref{eq:BHu}) can be expressed in terms of the integrals as
\begin{equation}\label{eq:T3}
\kappa_3^T= \frac{1}{2}\mathrm{I_A}-\frac{3}{2}\mathrm{I_B}+\frac{1}{8}\mathrm{I_C}+\frac{1}{8}\mathrm{I_D}
\end{equation}
with
\begin{eqnarray}
\mathrm{I_A}&=& \int_0^{\infty} f^3(x)\left(K_{01}(x, x)+K_{10}(x, x)\right)~\mathrm{d} x\\
\mathrm{I_B}&=&\mathrm{I^{(1)}_B}+\mathrm{I^{(2)}_B}+\mathrm{I^{(3)}_B}+\mathrm{I^{(4)}_B}\\ 
\mathrm{I_C}&=& \int_0^{\infty}\int_0^{\infty}\int_0^{\infty} f(x)f(y)f(z)~h_{\text{C}}~\mathrm{d} x \mathrm{d} y\mathrm{d} z\\
\mathrm{I_D}&=&\int_0^{\infty}\int_0^{\infty}\int_0^{\infty} f(x)f(y)f(z)~h_{\text{D}}~\mathrm{d} x \mathrm{d} y\mathrm{d} z,
\end{eqnarray}
where
\begin{eqnarray}
\mathrm{I^{(1)}_B}&=&\int_0^{\infty} \int_0^{\infty} f^2(x)f(y)K_{01}(x, y) K_{01}(y, x)\mathrm{d} x \mathrm{d} y\\
\mathrm{I^{(2)}_B}&=&\int_0^{\infty} \int_0^{\infty} f^2(x)f(y)K_{10}(x, y) K_{10}(y, x)\mathrm{d} x\mathrm{d} y\\
\mathrm{I^{(3)}_B}&=&\int_0^{\infty} \int_0^{\infty} f^2(x)f(y)K_{00}(x, y) K_{11}(x, y)\mathrm{d} x \mathrm{d} y\\
\mathrm{I^{(4)}_B}&=&\int_0^{\infty} \int_0^{\infty} f^2(x)f(y)K_{00}(x, y) K_{11}(x, y)\mathrm{d} x \mathrm{d} y.
\end{eqnarray}

\subsection{Calculation of cumulant integrals}\label{sec3.2}
Each integral in~(\ref{eq:T3}) can be similarly computed, which is outlined as follows. For $\mathrm{I}_{\mathrm{A}}$, consider 
\begin{equation}\label{eq:Aq}
A_{q}(t)=\int_{0}^{\infty}x^{\beta}(tx)^{2\alpha+1}H_{2\alpha+1-q}(tx)G_{q}(tx)\dd x,
\end{equation} where \begin{subequations}
\begin{eqnarray}
H_{q}(x)&=&G_{2,3}^{1,1}\left(\begin{array}{c}-m-2\alpha-1;m\\0;-q,-2\alpha-1\end{array}\Big|x\Big.\right)\\
G_{q}(x)&=&G_{2,3}^{2,1}\left(\begin{array}{c}-m-2\alpha-1;m\\0,-q;-2\alpha-1\end{array}\Big|x\Big.\right),
\end{eqnarray}
\end{subequations}
and take third order derivatives of $A_{q}(t)$ with respect to $\beta$ and evaluate for $q=\alpha, \alpha+1$, and $\beta=3$. By the definition of the kernel $h_1(x)$ in~\cite{LW21}, we then deduce that 
\begin{eqnarray}
\mathrm{I_A}&=&-\frac{2}{27}\left(A_{\alpha}+A_{\alpha+1}\right)\Big|_{\beta=3}+\frac{2}{9}\left(H_{\alpha}^{(1)}+H_{\alpha+1}^{(1)}\right) \nonumber\\
&&-\frac{1}{3}\left(H_{\alpha}^{(2)}+H_{\alpha+1}^{(2)}\right)+\frac{1}{3}\left(H_{\alpha}^{(3)}+H_{\alpha+1}^{(3)}\right),
\end{eqnarray}
where $A_q$ is the $t$ independent part of $A_q(t)$,
\begin{equation}\label{eq:Ibs0}
A_{q}(t)=t^{-\beta-1}A_{q},
\end{equation} and \begin{equation*}\label{eq:H123}
H_{q}^{(1)}=\frac{\dd A_{q}}{\dd\beta}\Big|_{\beta=3},~~~~~~H_{q}^{(2)}=\frac{\dd^{2} A_{q}}{\dd\beta^{2}}\Big|_{\beta=3},~~~~~~H_{q}^{(3)}=\frac{\dd^{3} A_{q}}{\dd\beta^{3}}\Big|_{\beta=3}.
\end{equation*}
To calculate $A_q(t)$, we utilize the finite sum representation for $H_q(tx)$ as in~\cite{Wei20BHA,Prudnikov86,Bertola14,FK16}, and apply the Mellin transform for Meijer G-function~\cite{Prudnikov86}
\begin{eqnarray}\label{eq:iMG}
&&\int_{0}^{\infty}x^{s-1}G_{p,q}^{m,n}\!\left(\begin{array}{c} a_{1},\ldots,a_{n}; a_{n+1},\ldots,a_{p} \\ b_{1},\ldots,b_{m}; b_{m+1},\ldots,b_{q} \end{array}\Big|\eta x\Big.\right)\dd x \nonumber \\
&=&\frac{\eta^{-s}\prod_{j=1}^m\Gamma\left(b_j+s\right)\prod_{j=1}^{n}\Gamma\left(1-a_j-s\right)}{\prod_{j=n+1}^{p}\Gamma\left(a_{j}+s\right)\prod_{j=m+1}^q\Gamma\left(1-b_j-s\right)}
\end{eqnarray} to $G_q(tx)$. 

Similarly as in~\cite{LW21}, by the symmetry we divide $\mathrm{I_C}$ into eight integrals
\begin{equation}\label{eq:C}
\mathrm{I_C}=4\left(2\mathrm{I^{(1)}_C}+\mathrm{I^{(2)}_C}-2\mathrm{I^{(3)}_C}-\mathrm{I^{(4)}_C}+2\mathrm{I^{(5)}_C}+\mathrm{I^{(6)}_C}-2\mathrm{I^{(7)}_C}-\mathrm{I^{(8)}_C}\right),
\end{equation}
where
\begin{eqnarray}
\mathrm{I^{(1)}_C} &=& \int_{0}^{\infty}\!\!\int_{0}^{\infty}\!\!\int_{0}^{\infty}\!\!f(x)f(y)f(z)K_{00}(x,y)K_{01}(y,z)K_{11}(x,z)\dd x\dd y\dd z\\
\mathrm{I^{(2)}_C} &=& \int_{0}^{\infty}\!\!\int_{0}^{\infty}\!\!\int_{0}^{\infty}\!\!f(x)f(y)f(z)K_{00}(y,z)K_{01}(y,x)K_{11}(x,z)\dd x\dd y\dd z\\
\mathrm{I^{(3)}_C} &=& \int_{0}^{\infty}\!\!\int_{0}^{\infty}\!\!\int_{0}^{\infty}\!\!f(x)f(y)f(z)K_{00}(x,y)K_{01}(x,z)K_{11}(y,z)\dd x\dd y\dd z\\
\mathrm{I^{(4)}_C} &=& \int_{0}^{\infty}\!\!\int_{0}^{\infty}\!\!\int_{0}^{\infty}\!\!f(x)f(y)f(z)K_{00}(y,z)K_{01}(z,x)K_{11}(x,y)\dd x\dd y\dd z\\
\mathrm{I^{(5)}_C} &=& \int_{0}^{\infty}\!\!\int_{0}^{\infty}\!\!\int_{0}^{\infty}\!\!f(x)f(y)f(z)K_{00}(x,y)K_{10}(z,x)K_{11}(z,y)\dd x\dd y\dd z\\
\mathrm{I^{(6)}_C} &=& \int_{0}^{\infty}\!\!\int_{0}^{\infty}\!\!\int_{0}^{\infty}\!\!f(x)f(y)f(z)K_{00}(y,z)K_{10}(x,z)K_{11}(y,x)\dd x\dd y\dd z\\
\mathrm{I^{(7)}_C} &=& \int_{0}^{\infty}\!\!\int_{0}^{\infty}\!\!\int_{0}^{\infty}\!\!f(x)f(y)f(z)K_{00}(x,y)K_{10}(z,y)K_{11}(z,x)\dd x\dd y\dd z\\
\mathrm{I^{(8)}_C} &=& \int_{0}^{\infty}\!\!\int_{0}^{\infty}\!\!\int_{0}^{\infty}\!\!f(x)f(y)f(z)K_{00}(y,z)K_{10}(x,y)K_{11}(z,x)\dd x\dd y\dd z.
\end{eqnarray}
The integral $\mathrm{I_D}$ is similarly divided into
\begin{equation}\label{eq:D}
\mathrm{I_D}=2\left(\mathrm{I^{(1)}_D}+3\mathrm{I^{(2)}_D}+3\mathrm{I^{(3)}_D}+\mathrm{I^{(4)}_D}\right),
\end{equation}
where
\begin{eqnarray}
\mathrm{I^{(1)}_D} &=& \int_{0}^{\infty}\!\!\int_{0}^{\infty}\!\!\int_{0}^{\infty}\!\!f(x)f(y)f(z)K_{01}(x,y)K_{01}(y,z)K_{01}(z,x)\dd x\dd y\dd z \\
\mathrm{I^{(2)}_D} &=& \int_{0}^{\infty}\!\!\int_{0}^{\infty}\!\!\int_{0}^{\infty}\!\!f(x)f(y)f(z)K_{01}(x,z)K_{01}(z,y)K_{10}(x,y)\dd x\dd y\dd z\\
\mathrm{I^{(3)}_D} &=& \int_{0}^{\infty}\!\!\int_{0}^{\infty}\!\!\int_{0}^{\infty}\!\!f(x)f(y)f(z)K_{01}(x,y)K_{10}(x,z)K_{10}(z,y)\dd x\dd y\dd z \\
\mathrm{I^{(4)}_D} &=& \int_{0}^{\infty}\!\!\int_{0}^{\infty}\!\!\int_{0}^{\infty}\!\!f(x)f(y)f(z)K_{10}(x,y)K_{10}(y,z)K_{10}(z,x)\dd x\dd y\dd z.
\end{eqnarray}
For integrals $\mathrm{I}^{(1)}_{\mathrm{B}}$, $\mathrm{I}^{(2)}_{\mathrm{B}}$, and $\mathrm{I_D}$ involving only correlation kernels of type $K_{01}$ and $K_{10}$ multiplying together, we utilize the kernel representation in~\cite{Bertola14} 
\begin{eqnarray}
K_{01}(x,y)&=&x^{2\alpha+1}\int_{0}^{1}t^{2\alpha+1}H_{\alpha}(ty)G_{\alpha+1}(tx)\dd t \label{eq:K01I} \\
K_{10}(x,y)&=&y^{2\alpha+1}\int_{0}^{1}t^{2\alpha+1}H_{\alpha+1}(tx)G_{\alpha}(ty)\dd t, \label{eq:K10I} 
\end{eqnarray}
and apply finite sum representations~\cite{Wei20BHA,Prudnikov86,Bertola14,FK16} of $G_{2,3}^{1,1}$ type Meijer G-functions, as well as the following identity of $G_{2,3}^{2,1}$ type Meijer G-functions~\cite{Wei20BH,Prudnikov86}
\begin{eqnarray}
&&\int_{0}^{1}x^{a-1}G_{p,q}^{m,n}\left(\begin{array}{c} a_{1},\ldots,a_{n}; a_{n+1},\ldots,a_{p} \\ b_{1},\ldots,b_{m}; b_{m+1},\ldots,b_{q} \end{array}\Big|\eta x\Big.\right)\dd x \nonumber\\
&=&G_{p+1,q+1}^{m,n+1}\left(\begin{array}{c}1-a,a_{1},\ldots,a_{n};a_{n+1},\ldots,a_{p}\\b_{1},\ldots,b_{m};b_{m+1},\ldots,b_{q},-a\end{array}\Big|\eta\Big.\right).
\end{eqnarray}
For example, for $\mathrm{I^{(1)}_D}$, consider integral 
\begin{eqnarray}
&&D_1\left(\beta_{1},\beta_{2},\beta_{3}\right)\nonumber\\
\!&=&\!\int_{0}^{\infty}\!\int_{0}^{\infty}\!\!\int_{0}^{\infty}x^{\beta_{1}}y^{\beta_{2}}z^{\beta_{3}}~K_{01}(x,y)K_{01}(y,z)K_{01}(z,x)\dd x\dd y\dd z \nonumber \\
&=&\!\sum _{i=0}^{m-1} \!\sum _{j=0}^{m-1} \!\sum _{k=0}^{m-1} \!\frac{(-1)^{i+j+k} \prod_{\substack{l=i,j,k}}\Gamma (l+m+2 \alpha +2)) }{\displaystyle{\prod\limits_{l=i,j,k}}(\Gamma (l+1) \Gamma (l+\alpha +1) \Gamma (l+2 \alpha +2) \Gamma (m-l)) }\nonumber\\
  && \! \times\!\!\int_0^{\infty } y^{i+2 \alpha +\beta_2} G_{3,4}^{2,2}\left(
\begin{array}{c}
 -j-2 \alpha -1,-m-2 \alpha -1,m \\
 0,-\alpha -1,-2 \alpha -1,-j-2 \alpha -2 \\
\end{array}
\Big| y \right) \, \!\mathrm{d}y\nonumber\\
&&\!\times\!\!\int_0^{\infty } x^{k+2 \alpha+\beta_1} G_{3,4}^{2,2}\left(
\begin{array}{c}
 -i-2 \alpha -1,-m-2 \alpha -1,m \\
 0,-\alpha -1,-2 \alpha -1,-i-2 \alpha -2 \\
\end{array}
\Big| x\right) \, \!\mathrm{d}x\nonumber\\
&&\!\times\!\!\int_0^{\infty } z^{j+2 \alpha +\beta_3} G_{3,4}^{2,2}\left(
\begin{array}{c}
 -k-2 \alpha -1,-m-2 \alpha -1,m \\
 0,-\alpha -1,-2 \alpha -1,-k-2 \alpha -2 \\
\end{array}
\Big|z \right) \, \!\mathrm{d}z,
\end{eqnarray}
and apply Mellin transform of Meijer G-function~(\ref{eq:iMG}) to the integrals above, $\mathrm{I^{(1)}_D}$ becomes
\begin{equation}
\mathrm{I^{(1)}_D}=\frac{\partial^{3}}{\partial\beta_{1}\partial\beta_{2}\partial\beta_{3}}D_{1}\left(\beta_{1},\beta_{2},\beta_3\right)\Big|_{\beta_{1}=1,\beta_{2}=1,\beta_{3}=1}.
\end{equation}
The other $\mathrm{I_D}$ integrals and $\mathrm{I}^{(1)}_{\mathrm{B}}$, $\mathrm{I}^{(2)}_{\mathrm{B}}$ are calculated similarly.

For integrals $\mathrm{I}_{\mathrm{B}}^{(3)}$ and $\mathrm{I}_{\mathrm{B}}^{(4)}$ involving only correlation kernels of type $K_{00}$ and $K_{11}$, we use the kernel representation in~\cite{FK16,Bertola14}
\begin{eqnarray}
K_{00}(x,y)&=&\sum_{k=0}^{m-1}\frac{1}{h_k}p_{k}(x)q_{k}(y) \label{eq:K00} \\
K_{11}(x,y)&=&x^{\alpha}\!y^{\alpha+1}e^{-x-y}\sum_{k=0}^{m-1}\frac{1}{h_k}P_{k}(-y)Q_{k}(-x) -W(x,y), \label{eq:K11}
\end{eqnarray}
where 
\begin{equation}\label{eq:w}
W(x,y)=\frac{x^{\alpha}y^{\alpha+1}e^{-x-y}}{x+y}
\end{equation}
is the weight function of the Cauchy-Laguerre biorthogonal polynomials $p_{k}(x)$ and $q_{l}(y)$,
\begin{equation}\label{eq:oc}
\int_{0}^{\infty}\int_{0}^{\infty}p_{k}(x)q_{l}(y)W(x,y)\dd x\dd y=h_k\delta_{kl}
\end{equation}
with the normalizing constant $h_k$. The Cauchy transforms of $p_{k}(x)$ and $q_{k}(y)$ are respectively
\begin{subequations}
\begin{eqnarray}\label{eq:PQ}
P_{k}(x)&=&\int_{0}^{\infty}\frac{v^{\alpha}\e^{-v}}{x-v}p_{k}(v)\dd v \\
Q_{k}(y)&=&\int_{0}^{\infty}\frac{w^{\alpha+1}\e^{-w}}{y-w}q_{k}(w)\dd w,
\end{eqnarray}
\end{subequations}
where their representations via Meijer G-functions are~\cite{Bertola10,Bertola14,Forrester}
\begin{subequations}\label{eq:kerM}
\begin{align}
p_{k}(x)=&(-1)^k \frac{\Gamma(k+1)\Gamma(2\alpha+k+2)\Gamma(\alpha+k+1)}{\Gamma(2\alpha+2k+2)}\nonumber\\
&\times G_{2,3}^{1,1}\left(\left.
\begin{array}{c}
-2\alpha-k-1;~k+1\\
0;~-\alpha,\,-2\alpha-1\end{array}
\right|x
\right)\\
q_k(y)=&(-1)^k \frac{\Gamma(k+1)\Gamma(2\alpha+k+2)\Gamma(\alpha+k+2)}{\Gamma(2\alpha+2k+2)}\nonumber\\
&\times G_{2,3}^{1,1}\left(\left.
\begin{array}{c}
-2\alpha-k-1;~k+1\\
0;-\alpha-1,\,-2\alpha-1
\end{array}
\right|y
\right)\\
P_k(x)=&(-1)^{k+1}\frac{2\alpha+2k}{\Gamma(k)\Gamma(\alpha+k)}G_{2,3}^{3,1}\left(\!
\left.\begin{array}{c}
-k;\,k+2\alpha\\
-1,\alpha-1,2\alpha;
\end{array}\right|-x\!
\right)\\
Q_k(y)=&(-1)^{k+1}\frac{2\alpha+2k}{\Gamma(k)\Gamma(\alpha+k+1)}G_{2,3}^{3,1}\left(\!
\left.\begin{array}{c}
-k;\,k+2\alpha\\
-1,\alpha,2\alpha;
\end{array}\right|-y\!
\right).
\end{align}
\end{subequations}

We now consider the integral 
\begin{equation}
B_{3,4}\left(\beta_{1},\beta_{2}\right)=\int_{0}^{\infty}\!\!\int_{0}^{\infty}x^{\beta_{1}}y^{\beta_{2}}~K_{00}(x,y)K_{11}(x,y)\dd x\dd y,
\end{equation}
and utilize the Mellin transform of Meijer G-function (\ref{eq:iMG}) to arrive at 
\begin{eqnarray}
&& B_{3,4}\left(\beta_{1},\beta_{2}\right)\label{eq:di}\nonumber\\
\!\!\!\!\!&=&\!\!\!\sum_{j=0}^{m-1}\!\sum_{k=0}^{m-1}\!\sum_{i=0}^{j}\!\sum_{s=0}^{j}\!\frac{4(-1)^{i+s}(\alpha\!+\!j\!+\!1)(\alpha\!+\!k\!+\!1)\Gamma(2\alpha\!+\!i\!+\!j\!+\!2)\Gamma(2\alpha\!+\!j\!+\!s\!+\!2)}{i!s!\Gamma(\alpha+i+1)\Gamma(2\alpha+i+2)\Gamma(\alpha+s+2)\Gamma(2\alpha+s+2)}\nonumber\\
\!\!&&\!\!\times\frac{\Gamma\left(\beta_1+i+1\right)\Gamma\left(\alpha+\beta_1+i+1\right)\Gamma\left(2\alpha+\beta_1+i+2\right)}{\Gamma\left(\beta_1+i-k+1\right)\Gamma\left(2\alpha+\beta_1+i+k+3\right)\Gamma(j-i+1)}\nonumber\\
\!\!&&\!\!\times\frac{\Gamma\left(\beta_2\!+\!s\!+\!1\right)\Gamma\left(\alpha\!+\!\beta_2\!+\!s\!+\!2\right)\Gamma\left(2\alpha\!+\!\beta_2\!+s\!+\!2\right)}{\Gamma\left(\beta_2\!-\!k\!+\!s\!+1\right)\Gamma\left(2\alpha\!+\!\beta_2\!+\!k\!+\!s\!+3\right)\Gamma(j\!-\!s\!+\!1)}-\!\sum_{j=0}^{m-1}\!\sum_{i=0}^{j}\!\sum_{k=0}^{j}\!\frac{2(-1)^{i+k}}{i!k!}\nonumber\\
\!\!&&\!\!\times\frac{(\alpha+j+1)\Gamma(2\alpha+i+j+2)\Gamma(2\alpha+j+k+2)}{\Gamma(j\!-\!i\!+\!1)\Gamma(j\!-\!k\!+\!1)\Gamma(\alpha\!+\!i\!+\!1)\Gamma(2\alpha\!+\!i\!+\!2)\Gamma(\alpha\!+\!k\!+\!2)\Gamma(2\alpha\!+\!k\!+\!2)}\nonumber\\
\!\!&&\!\!\times\frac{\Gamma\left(\alpha+\beta_1+i+1\right)\Gamma\left(\alpha+\beta_2+k+2\right)}{2\alpha+\beta_1+\beta_2+i+k+2}\label{eq:4sums}.
\end{eqnarray}
Note that, one piece of the summation in the above quadruple sum can be evaluated through Lemma 4.1 in \cite{Bertola14}, cf. \cite{Wei20BH}. We calculate $\mathrm{I^{(3)}_B}$ and $\mathrm{I^{(4)}_B}$ by taking different orders of derivatives of $\beta_{1}$ and $\beta_{2}$ respectively, 
\begin{eqnarray}
\mathrm{I^{(3)}_B}&=&\frac{\partial^{3}}{\partial\beta_{1}^2\partial\beta_{2}}B_{3,4}\left(\beta_{1},\beta_{2}\right)\Big|_{\beta_{1}=2,\beta_{2}=1} \\
\mathrm{I^{(4)}_B}&=&\frac{\partial^{3}}{\partial\beta_{1}\partial\beta_{2}^2}B_{3,4}\left(\beta_{1},\beta_{2}\right)\Big|_{\beta_{1}=1,\beta_{2}=2}.
\end{eqnarray}
 
For $\mathrm{I_C}$ integrals involving kernels of type $K_{01}$ or $K_{10}$, as well as $K_{00}$ and $K_{11}$, we use a combination of the above two methods to the two kinds of kernels. In taking derivatives with respect to the parameter $\beta$ in the summation form of the integrals, we exchange the order of taking derivatives and the finite summation, and end up with summations of rational functions involving Gamma and polygamma functions.

Finally, by resolving the indeterminacy of polygamma functions with negative arguments in the limit $\epsilon\to0$ using~\cite{Wei20,Wei20BH},
\begin{subequations}\label{eq:pgna}
\begin{eqnarray}
\Gamma(-l+\epsilon)& =&\frac{(-1)^l}{l!\epsilon}\left(1+\psi_0(l+1) \epsilon+o\left(\epsilon^2\right)\right) \\
\psi_0(-l+\epsilon)&=&-\frac{1}{\epsilon}+\psi_0(l+1)+\left(2 \psi_1(1)-\psi_1(l+1)\right)\epsilon \nonumber\\
&&+\frac{1}{2}\psi_2(l+1) \epsilon^2+o\left(\epsilon^3\right) \\
\psi_1(-l+\epsilon)&=&\frac{1}{\epsilon^2}-\psi_1(l+1)+\psi_1(1)+\zeta(2)+o(\epsilon) \\
\psi_2(-l+\epsilon)&=&-\frac{2}{\epsilon^3}+\psi_2(l+1)+\psi_2(1)+2 \zeta(3)+o(\epsilon),
\end{eqnarray}
\end{subequations}
all integrals from $\mathrm{I_A}$ to $\mathrm{I_D}$ in~(\ref{eq:T3}) are now computed into nested summations. The remaining task is to simplify the summations into expressions involving single sums of anomalies as summarized in Table~\ref{tab:exist} and Table~\ref{tab:basis}. Summation identities utilized in the simplification are listed in Appendix~\ref{secA1}. The simplification of summations is a case-by-case task, which, in the current case of third cumulant, is much more tedious as compared to that of the mean~\cite{Wei20BHA} and the variance~\cite{Wei20BH}. In particular, the number of anomalies increases rapidly as seen by comparing Table~\ref{tab:basis} to Table~\ref{tab:exist}. This calls for tailor-made summation identities to transform anomalies into each other before adding up to the closed-form result~(\ref{eq:main}). The derivation of the new summations identities is presented in Section~\ref{sec3.3} below. 

\subsection{Derivation of new summation identities}\label{sec3.3}
Finding the desired summation identities to relate anomalies relies on the re-summation technique, which is systematically discussed in~\cite{huang_entropy_2023} but has been utilized in prior works~\cite{HW22,huang_entropy_2023,HWISIT,HWC21,Wei17,Wei19,Wei20,Wei20BHA,Wei20BH,Wei22}. The main idea is to identify alternative forms of a finite sum by iterating a suitable recurrence relation. Using re-summation for anomalies with both polygamma functions of positive and negative summation variables on the numerator, i.e., $\Omega_{12}^{(a)}$ to $\Omega_{14}^{(a)}$ in Table~\ref{tab:basis}, the derivation requires case-wise choices, and re-summing certain index would again give out single-sum anomalies. Therefore, the re-summation needs to be applied twice for such anomalies. 

The identity~(\hyperlink{17}{\ref*{17}}) is a reasonably representative one, the derivation of which using the re-summation technique is discussed below. Consider the re-summation function $G(m)$,
\begin{equation}
G(m)=\sum\limits_{\smash{k=1}}^m \frac{\psi_0(k)\psi_0(b+m+1-k)}{k},
\end{equation} 
where $b=a-m$, one has
\begin{eqnarray}
G(i)-G(i-1)&=&\frac{\psi _0(i) \psi _0(b+1)}{i}+\sum\limits_{\smash{k=1}}^{i-1} \frac{\psi _0(k)}{k(i+b-k)}.\label{eq:g11}
\end{eqnarray}
Summing over $i$ on both side of~(\ref{eq:g11}) above yields
\begin{eqnarray}
G(m)&=& \psi _0(b+1)\sum\limits_{\smash{i=1}}^m\frac{\psi _0(i)}{i}+\sum\limits_{\smash{i=1}}^m\sum\limits_{\smash{k=1}}^{i-1}\frac{\psi_0 (k)}{k (i+b)}+\sum\limits_{\smash{i=1}}^m\sum\limits_{\smash{k=1}}^{i-1}\frac{\psi_0 (k)}{(i+b) (i+b-k)}\nonumber\\
&=&\frac{1}{2} \left(\psi_0^2 (m+1)+ \psi _1(m+1)- \psi_0^2 (1)- \psi _1(1)\right)\psi _0(b+1)\nonumber\\
&&+\sum\limits_{\smash{i=1}}^m\frac{1}{2(i+b)} \left( \psi_0^2 (i)+ \psi _1(i)- \psi_0^2 (1)- \psi _1(1)\right)\nonumber\\
&&+\sum\limits_{\smash{i=1}}^m\frac{1}{i+b}\bigg(\frac{1}{2} \big(\left(\psi _0(i+b)-\psi _0(b)\right)^2+2 \psi _0(i) \left(\psi _0(i+b)+\psi _0(b)\right.\nonumber\\
&&\left.-\psi _0(b+1)\right)+\psi _1(i+b)-2 \psi _0(1) \psi _0(b)-\psi _1(b)\big)\bigg)\nonumber\\
&&-\sum\limits_{\smash{i=1}}^m\frac{1}{i+b}\sum _{k=1}^{i-1} \frac{\psi _0(k+b)}{k},
\end{eqnarray}
where~(\hyperlink{4}{\ref*{4}}) has also been utilized. The resulting summation 
\begin{equation}
\sum\limits_{\smash{i=1}}^m\frac{\psi_0(i)\psi _0(i+b)}{i+b}
\end{equation} is evaluated via~(\hyperlink{21}{\ref*{21}}), and the last term is calculated by changing summation order as
\begin{eqnarray}
   \sum\limits_{\smash{i=1}}^m \sum\limits_{\smash{k=1}}^{i-1}\frac{\psi _0(k+b)}{k(i+b)} &=&  \sum\limits_{\smash{k=1}}^{m-1} \sum\limits_{\smash{i=k+1}}^{m}\frac{\psi _0(k+b)}{k(i+b)}\nonumber\\
   &=& \psi_0(b+m+1)\sum\limits_{\smash{k=1}}^{m-1}\frac{\psi _0(k+b)}{k}\nonumber\\
   &&-\sum\limits_{\smash{k=1}}^{m-1}\frac{\psi _0(k+b)\psi _0(k+b+1)}{k}.
\end{eqnarray}
Shifting the summation index as needed before inserting back $b=a-m$, one obtains~(\hyperlink{17}{\ref*{17}}). 

Similar tricks have been used in deriving~(\hyperlink{3}{\ref*{3}}),~(\hyperlink{4}{\ref*{4}}),~(\hyperlink{10}{\ref*{10}})--(\hyperlink{16}{\ref*{16}}),~(\hyperlink{18}{\ref*{18}}), and~(\hyperlink{19}{\ref*{19}}), where the corresponding re-summation functions are listed in Appendix \ref{secA2}. Note that, in the derivation of~(\hyperlink{18}{\ref*{18}}) via the re-summation~(\ref{eq:f18}), we make use of~(\hyperlink{3}{\ref*{3}}) for 
\begin{equation}
\sum\limits_{k=1}^{i-1}\frac{\psi_0(k+b)}{i-k} 
\end{equation}
as
\begin{eqnarray}
\sum _{i=1}^{m}   \sum _{k=1}^{i-1} \frac{\psi _0(k+b)}{k(i-k)}
&=& \sum _{i=1}^{m}   \sum _{k=1}^{i-1} \frac{\psi_0(k+b)}{ik}+ \sum _{i=1}^{m}   \sum _{k=1}^{i-1} \frac{\psi _0(k+b)}{i(i-k)}\nonumber\\
&=& \sum _{i=1}^{m}   \sum _{k=1}^{i-1}\frac{1}{i} \frac{\psi _0(k+b)}{k}+ \sum _{i=1}^{m}   \sum _{k=1}^{i-1}\frac{1}{i} \frac{\psi _0(i+b-k)}{k}\nonumber\\
&=& \sum _{i=1}^{m}   \frac{1}{i} \sum _{k=1}^{i-1}  \frac{\psi _0(k+b)}{k}- \sum _{i=1}^{m}\frac{1}{i}   \sum _{k=1}^{i-1}  \frac{\psi _0(k+b)}{k}\nonumber\\
&&+\frac{1}{2}\sum _{i=1}^{m} \frac{1}{i}\big(\!-2 \psi _0(i+b) \big(\psi _0(b)-\psi _0(i)+\psi _0(1)\big)\nonumber\\
&&+\big(\psi _0(b)+2 \psi _0(i)-2 \psi _0(1)\big) \psi _0(b)\nonumber\\
&&-\psi _1(b)+\psi _0^2(i+b)+\psi _1(i+b)\big),
\end{eqnarray}
and make use of~(\hyperlink{20}{\ref*{20}}) for a resulting single-sum
\begin{equation}
\sum\limits_{i=1}^{m}\frac{\psi_0(i)\psi_0(i+b)}{i}.
\end{equation} 

Note also that the re-summation~(\ref{eq:f19}) used for deriving~(\hyperlink{19}{\ref*{19}}) is slightly different as the re-summation index is on the denominator, where we consider 
\begin{equation}
G(m)=\sum_{k=1}^{m}\frac{\psi_0(k)\psi_0(k+b)}{m+1-k},
\end{equation}
so that 
\begin{eqnarray}\label{eq:19}
&& G(i)-G(i-1)\nonumber\\
   &=&\sum _{k=1}^{i} \frac{\psi_0 (k) \psi_0 (k+b)}{i+1-k}
     -\sum _{k=1}^{i-1} \frac{\psi_0 (k) \psi_0 (k+b)}{i-k}\nonumber\\
        &=&\sum _{k=1}^{i-1} \frac{\psi_0 (k+1) \psi_0 (k+b+1)}{i-k}-\sum _{k=1}^{i-1} \frac{\psi_0 (k) \psi_0 (k+b)}{i-k}+\frac{\psi_0 (1) \psi_0 (b+1)}{i}\nonumber\\
        &=&\frac{1}{i}\left(\sum _{k=1}^{i-1}\frac{\psi_0(k+b)}{k}+\sum _{k=1}^{i-1}\frac{\psi_0(k+b)}{i-k}\right)\nonumber\\
        &&+\frac{1}{i+b}\left(\sum _{k=1}^{i-1}\frac{\psi_0(k)}{k+b}+\sum _{k=1}^{i-1}\frac{\psi_0(k)}{i-k}\right)+\frac{\psi_0 (1) \psi_0 (b+1)}{i}.
\end{eqnarray}
The resulting single sums 
\begin{eqnarray}
&&\sum\limits_{k=1}^{i-1}\frac{\psi_0(k+b)}{i-k} \\
&&\sum\limits_{k=1}^{i-1}\frac{\psi_0(k)}{k+b}  \\
&&\sum\limits_{k=1}^{i-1}\frac{\psi_0(k)}{i-k}
\end{eqnarray} 
are transformed to 
\begin{equation}
\sum\limits_{k=1}^{i-1}\frac{\psi_0(k+b)}{k}
\end{equation} via~(\hyperlink{3}{\ref*{3}}),~(\hyperlink{0}{\ref*{0}}), and (\ref{1}), respectively. We then calculate $G(m)$ by summing over~(\ref{eq:19}) as 
\begin{equation}
G(m)=\sum_{i=1}^{m}\left(G(i)-G(i-1)\right),
\end{equation}
where the sum over $i$ is evaluated first in the resulting double-sums.

In addition to the re-summation technique, relations of polygamma functions that differ by positive integer arguments~(\ref{eq:psisre}) such as
\begin{subequations}\label{shift}
\begin{eqnarray}
\psi_0(l+n)&=&\psi_0(l)+\sum_{i=0}^{n-1}\frac{1}{l+i}\label{shift0}\\
\psi_1(l+n)&=&\psi_1(l)-\sum_{i=0}^{n-1}\frac{1}{(l+i)^2}
\end{eqnarray}
\end{subequations}
are also useful to obtain some of the new summation identities. For anomalies with two polygamma functions on the numerator and positive summation variable arguments, we now derive for distinct positive integers $a$, $b$ and $c$, the summation identity~(\ref{formula}) relating the three anomalies by expanding and rearranging the summation. Specifically, we write by~(\ref{shift0}),
\begin{equation}
\sum _{k=1}^{m}\frac{\psi_0(k+b)\psi_0(k+c)}{k+a}= \sum \limits_{k=1}^{m}\frac{\left(\psi_0(b)+\sum\limits_{i=0}^{k-1}\frac{1}{b+i}\right)\left(\psi_0(c)+\sum\limits_{j=0}^{k-1}\frac{1}{c+j}\right)}{k+a}
\end{equation}
before changing the summation order in resulting triple sum. The result~(\ref{formula}) can be viewed as a generalized three-term version of the two-term relation (\hyperlink{cb}{\ref*{cb}}). We then obtain (\hyperlink{20}{\ref*{20}}), (\hyperlink{21}{\ref*{21}}) by setting $a=1$, $c=1$, or $a=1$, $c=b+1$, respectively, in (\ref{formula}) with the needed shift of arguments via (\ref{shift0}).

Summation identities for certain anomalies involving trigamma functions adding up to closed forms such as (\ref{24}) and (\ref{25}) are derived by taking partial derivatives of (\ref{23}). In particular, the identity (\ref{24}) is obtained by taking partial derivative with respect to $b$ in (\ref{23}), evaluate at $a=0$, $b=2\alpha$ and utilize~(\hyperlink{cb}{\ref*{cb}}), whereas the identity (\ref{25}) is obtained by taking partial derivative with respect to $a$ in (\ref{23}), evaluate at $a=0$, $b=2\alpha$ and utilize~(\hyperlink{cb}{\ref*{cb}}) and (\ref{24}).\\

With the new anomalies in Table \ref{tab:basis} being processed using the identities in Appendix \ref{secA1} derived above, all anomalies cancel completely when collecting them in the third cumulant expression~(\ref{eq:T3}) that leads to
\begin{eqnarray}\label{kappa3T}
\!\!\!\!\!\!\!\!\!\kappa_3^{T} &=&m(2n-m)\left(b_1\psi_2\!\left(n+\frac{1}{2}\right)+b_2\psi_0\!\left(n+\frac{1}{2}\right)\psi_1\!\left(n+\frac{1}{2}\right)\right. \nonumber \\
&&+b_3\psi_1\!\left(n+\frac{1}{2}\right)\!+\psi_0^3\!\left(n+\frac{1}{2}\right)\!\left.+\frac{9}{2}\psi_0^2\!\left(n+\frac{1}{2}\right)\!+3 \psi_0\!\left(n+\frac{1}{2}\right)\!\right)\!,
\end{eqnarray}
where\begin{eqnarray*}
b_1&=&\frac{-4 m^2+8 m n+4 n^2+7}{8}\\
b_2&=&\frac{3\big(\!-m^2+2 m n+4 n^2+1\big)}{4n}\\
b_3&=&\frac{-m^4-16 m^2 n^2+4 m^2 n+5 m^2+24 m n^3-10 m n+24 n^4+10 n^2
-4}{2n(2n-1)(2n+1)}.
\end{eqnarray*}
Finally, inserting the obtained cumulant formulas~(\ref{eq:vNm}),~(\ref{eq:vNv}),~(\ref{kappa3T}) into the result~(\ref{mc}) while keeping in mind the moment-cumulant relation~(\ref{eq:m2c}), the claimed main result~(\ref{eq:main}) in Proposition~\ref{p:1} is established.

\appendices
\counterwithin{equation}{section}
\section{Summation Identities}\label{secA1}
We summarize identities utilized in the simplification of nested summations resulting from the integrals in Section~\ref{sec3.2}, where the identities~(\ref{1})--(\hyperlink{cb}{\ref*{cb}}), (\ref{23}) are known in the literature~\cite{Wei20BH,huang_entropy_2023}, and the rest are new. 

Let $ m$ be a positive integer, we have $a \geq m$ for~(\hyperlink{3}{\ref*{3}}), $b \geq 0$, $c \geq 0$ for~(\hyperlink{0}{\ref*{0}}),~(\hyperlink{cb}{\ref*{cb}}), $b > 0$ for~(\hyperlink{7}{\ref*{7}}),~(\hyperlink{8}{\ref*{8}}),~(\hyperlink{20}{\ref*{20}}), (\hyperlink{21}{\ref*{21}}), (\ref{23}), $a > m$ for~(\hyperlink{10}{\ref*{10}})--(\hyperlink{19}{\ref*{19}}), and $a$, $b$, $c$ being three distinct positive integers in (\ref{formula}). The identities~(\ref{1})--(\hyperlink{21}{\ref*{21}}) transfer various complicated anomalies into basic ones, whereas the identities (\ref{formula})--(\ref{25}) are examples of anomalies add up to closed-form expressions.
\makeatletter
\tagsleft@true
\begin{align}
~~~~\!\! \sum _{\smash{k=1}}^m \frac{\psi _0(k)}{m+1-k}=\psi _0^2(m+1)-\psi _0(1) \psi _0(m+1)+\psi _1(m+1)-\psi _1(1)\!\!\!\!\!\!\!\!\!\!\!\label{1}
\end{align}
\begin{align}
~~~~\!\!\sum _{\smash{k=1}}^m \frac{\psi _0(m+1-k)}{k}=\psi _0^2(m+1)-\psi _0(1) \psi _0(m+1)+\psi_1(m+1)-\psi _1(1) \!\!\!\!\!\!\!\!\!\!\!\!\!\!\!\!\!\!\!\!\!\!\label{2}
\end{align}
\tagsleft@false
\makeatother
\hypertarget{0}{}\begin{dgroup}\label{0}
\begin{dmath*}
\sum_{\smash{k=1}}^{m}\frac{\psi_{0}(k+b)}{k+c}=-\sum_{k=1}^{m}\frac{\psi_{0}(k+c)}{k+b}+\psi_{0}(m+c+1)\psi_{0}(m+b+1)-\psi_{0}(c+1)\psi_{0}(c+1)
+\frac{1}{c-b}\big(\psi_{0}(m+a+1)-\psi_{0}(m+b+1)-\psi_{0}(c+1)+\psi_{0}(b+1)\big) \qquad\qquad\qquad\qquad\qquad\qquad
\end{dmath*}
\end{dgroup}
\hypertarget{3}{}\begin{dgroup}\label{3}
\begin{dmath*}
\sum _{\smash{k=1}}^m \frac{\psi _0(a+1-k)}{k}=
-\sum _{\smash{k=1}}^m \frac{\psi _0(k+a-m)}{k}+\frac{1}{2} \big(-2 \psi _0(a+1) \big(\psi _0(a-m)-\psi _0(m+1)+\psi _0(1)\big)+\big(\psi _0(a-m)+2 \psi _0(m+1)-2 \psi _0(1)\big) \psi _0(a-m)-\psi _1(a-m)+\psi _0^2(a+1)+\psi _1(a+1)\big)\\
\end{dmath*}
\end{dgroup}
\hypertarget{4}{}\begin{dgroup}\label{4}
\begin{dmath*}
\sum _{\smash{k=1}}^m \frac{\psi _0(k)}{a+1-k}=-\sum _{\smash{k=1}}^m \frac{\psi _0(k+a-m)}{k}+\frac{1}{2} \big(\big(\psi _0(a+1)-\psi _0(a-m)\big)^2+2 \psi _0(m+1) \\
\times\big(-\psi _0(a-m+1)+\psi _0(a-m)+\psi _0(a+1)\big)-2 \psi _0(1) \psi _0(a-m)-\psi _1(a-m)+\psi _1(a+1)\big)\\
\end{dmath*}
\end{dgroup}
\hypertarget{sp1}{}\begin{dgroup}\label{sp1}
\begin{dmath*}
\sum_{\smash{k=1}}^m  \frac{\psi _0(k)}{k^2}=\sum_{\smash{k=1}}^m \frac{\psi _1(k)}{k}-\psi _0(m+1) \psi _1(m+1)-\frac{1}{2} \psi _2(m+1)+\psi _0(1) \psi _1(1)  +\frac{1}{2}\psi _2(1)\qquad\qquad\qquad\qquad\qquad\qquad\qquad\qquad\qquad\qquad\qquad\qquad\\
\\
\end{dmath*}
\end{dgroup}
\hypertarget{5}{}\begin{dgroup}\label{5}
\begin{dmath*}
\sum _{\smash{k=1}}^m \frac{\psi _0(m+1-k)}{k^2}=\sum _{\smash{k=1}}^m \frac{\psi _1(k)}{k}+\psi _0(1) \psi _1(m+1)+\psi _0(m+1) \big(\psi _1(1)-2 \psi _1(m+1)\big)-\psi _2(m+1)+\psi _2(1)\qquad\qquad\qquad\qquad\qquad\\
\\
\end{dmath*}
\end{dgroup}
\hypertarget{6}{}\begin{dgroup}\label{6}
\begin{dmath*}
\sum_{\smash{k=1}}^m \frac{\psi _0^2(m+1-k)}{k}=\sum_{\smash{k=1}}^m  \frac{\psi _1(k)}{k}+\psi _0^3(m+1)-\psi _0(1) \psi _0^2(m+1)-2 \psi _1(1) \psi _0(m+1)+\psi _1(m+1) \psi _0(m+1)+\psi _0(1) \psi _1(m+1)\quad\qquad\qquad\qquad\qquad\qquad\qquad\qquad\quad~~\\
\end{dmath*}
\end{dgroup}
\hypertarget{bb}{}\begin{dgroup}\label{bb}
\begin{dmath*}
\sum _{\smash{k=1}}^m \frac{\psi _0(k+b)}{(k+b)^2}= \sum _{\smash{k=1}}^m \frac{\psi _1(k+b)}{k+b} - \psi _0(b+m+1) \psi _1(b+m+1)-\frac{1}{2}\psi _2(b+m+1)+ \psi _0(b+1) \psi _1(b+1)+\frac{1}{2}\psi _2(b+1)\qquad\qquad\\ 
\end{dmath*}
\end{dgroup}
\hypertarget{7}{}\begin{dgroup}\label{7}
\begin{dmath*}
\sum _{\smash{k=1}}^m \frac{\psi _0(k+b) }{k^2}=\sum _{\smash{k=1}}^m \frac{\psi _1(k)}{k+b}-\frac{1}{b^2}\big(\psi _0(b+m+1)-\psi _0(b+1)-\psi _0(m+1)+\psi _0(1)\big)-\psi _1(m+1) \psi _0(b+m+1)-\frac{1}{b}\big(\psi _1(1)-\psi _1(m+1)\big)+\psi _1(1) \psi _0(b+1)\qquad\qquad\qquad\quad\qquad\quad\qquad\quad
\end{dmath*}
\end{dgroup}
\hypertarget{8}{}\begin{dgroup}\label{8}
\begin{dmath*}
\sum _{\smash{k=1}}^m \frac{\psi _0(k)}{(k+b)^2}=\sum _{k=1}^m \frac{\psi _1(k+b)}{k}+\frac{1}{b^2}\big(\psi _0(b+m+1)-\psi _0(b+1)-\psi _0(m+1)+\psi _0(1)\big)-\psi _0(m+1) \psi _1(b+m+1)-\frac{1}{b}\big(\psi _1(b+m+1)-\psi _1(b+1)\big)+\psi _0(1) \psi _1(b+1)\qquad\qquad\qquad\qquad\qquad\qquad\quad\quad\qquad\quad
\end{dmath*}
\end{dgroup}
\hypertarget{cb}{}\begin{dgroup}\label{cb}
\begin{dmath*}
\sum _{\smash{k=1}}^m \frac{\psi _0(k+b)}{(k+c)^2}=\sum _{\smash{k=1}}^m \frac{\psi _1(k+c)}{k+b}+\frac{1}{(c-b)^2}\big(\psi _0(c+m+1)-\psi _0(c+1)-\psi _0(b+m+1)+\psi _0(b+1)\big)+\frac{1}{c-b}\big(\psi _1(c+1)-\psi _1(c+m+1)\big)-\psi _1(c+m+1) \psi _0(b+m+1)+\psi _1(c+1) \psi _0(b+1)
\end{dmath*}
\end{dgroup}
\hypertarget{10}{}\begin{dgroup}\label{10}
\begin{dmath*}
\sum _{\smash{k=1}}^m \frac{\psi _0(k)}{(a+1-k)^2}=\sum _{\smash{k=1}}^m \frac{\psi _1(k+a-m)}{k}+\frac{1}{(a-m)^2}\big(-\psi _0(a-m+1)+\psi _0(a+1)-\psi _0(m+1)+\psi _0(1)\big)-\psi _1(a+1) \psi _0(m+1)+\psi _0(a+1) \big(\psi _1(a-m+1)-\psi _1(a+1)\big)+\frac{1}{a-m}\big(\psi _1(a-m+1)-\psi _1(a+1)\big)+\psi _0(a-m+1) \big(\psi _1(a+1)-\psi _1(a-m+1)\big)+\psi _0(1) \psi _1(a-m+1)+\frac{1}{2} \psi _2(a-m+1)-\frac{1}{2} \psi _2(a+1)
\end{dmath*}
\end{dgroup}
\hypertarget{11}{}\begin{dgroup}\label{11}
\begin{dmath*}
\sum _{\smash{k=1}}^m \frac{\psi _0^2(k)}{a+1-k}
=\sum _{\smash{k=1}}^m \bigg(\frac{\psi _0^2(k+a-m)}{k}+ \frac{\psi _0^2(k)}{k+a-m}+\frac{\psi _1(k+a-m)}{k+a-m}\bigg)+\Big(\frac{2}{a-m}-2 \psi _0(a+1)\Big)\sum _{\smash{k=1}}^m \frac{\psi _0(k+a-m)}{k}+\frac{1}{(a-m)^2}\big(\psi _0(a-m+1)-\psi _0(a+1)+\psi _0(m+1)-\psi _0(1)\big)+\frac{1}{a-m}\big(2 \psi _0(a+1) \big(-\psi _0(a-m+1)-\psi _0(m+1)+\psi _0(1)\big)+\psi _0^2(a-m+1)+\psi _0^2(a+1)\big)+\frac{1}{6} \big(-6 \psi _0^2(a+1) \big(\psi _0(a-m+1)-\psi _0(m+1)\big)-6 \psi _0(a+1) \big(-\psi _0^2(a-m+1)+2 \psi _0(1) \psi _0(a-m+1)+\psi _1(a-m+1)\big)-2 \psi _0^3(a-m+1)+6 \psi _0(1) \psi _0^2(a-m+1)-6 \psi _0(1) \psi _1(a-m+1)+6 \psi _0(a-m+1) \psi _1(a-m+1)+\psi _2(a-m+1)+2 \psi _0^3(a+1)+6 \psi _0(1) \psi _1(a+1)-\psi _2(a+1)\big)
\end{dmath*}
\end{dgroup}
\hypertarget{12}{}\begin{dgroup}\label{12}
\begin{dmath*}
\sum _{\smash{k=1}}^m \frac{\psi _1(a+1-k)}{k}=-\sum _{\smash{k=1}}^m \frac{\psi _1(k+a-m)}{k}+\frac{1}{(a-m)^2}\big(\psi _0(a-m+1)-\psi _0(a+1)+\psi _0(m+1)-\psi _0(1)\big)+\psi _1(a+1) \big(-\psi _0(a-m+1)+\psi _0(a+1)+\psi _0(m+1)-\psi _0(1)\big)+\frac{1}{a-m}\big(\psi _1(a+1)-\psi _1(a-m+1)\big)+\frac{1}{2} \big(2 \big(\psi _0(a-m+1)-\psi _0(a+1)+\psi _0(m+1)-\psi _0(1)\big) \psi _1(a-m+1)-\psi _2(a-m+1)+\psi _2(a+1)\big)\qquad\qquad\qquad\quad
\end{dmath*}
\end{dgroup}
\hypertarget{14}{}\begin{dgroup}\label{14}
\begin{dmath*}
\sum _{k=1}^m \frac{\psi _1(k)}{a+1-k}=\sum _{k=1}^m\bigg(- \frac{\psi _1(k+a-m)}{k}+ \frac{\psi _1(k)}{k+a-m}- \frac{\psi _1(k+a-m)}{k+a-m}\bigg)-\psi _1(m+1) \psi_0 (a-m+1)+\frac{1}{(a-m)^2}\big(\psi _0(a-m+1)-\psi _0(a+1)+\psi _0(m+1)-\psi _0(1)\big)+\frac{1}{a-m}\big(\psi _1(a+1)-\psi _1(a-m+1)\big)+\frac{1}{2} \big(-2 \psi _1(a+1) \big(\psi _0(a-m+1)-\psi _0(m+1)+\psi _0(1)\big)+2 \psi _1(m+1) \psi _0(a-m+1)-\psi _2(a-m+1)+\psi _2(a+1)\big)+\psi _0(a+1) \big(\psi _1(a+1)-\psi _1(1)\big)+\psi _1(1) \psi_0 (a+1)\qquad\qquad\qquad\qquad\qquad\qquad\qquad\qquad\qquad\quad
\end{dmath*}
\end{dgroup}
\hypertarget{15}{}\begin{dgroup}\label{15}
\begin{dmath*}
\sum _{\smash{k=1}}^m \frac{\psi _0(a+1-k)}{k^2}=\sum _{\smash{k=1}}^m\bigg( \frac{\psi _1(k+a-m)}{k}- \frac{\psi _1(k)}{k+a-m}+ \frac{\psi _1(k+a-m)}{k+a-m}\bigg)+\frac{1}{(a-m)^2}\big(-\psi _0(a-m+1)+\psi _0(a+1)-\psi _0(m+1)+\psi _0(1)\big)+\frac{1}{a-m}\big(\psi _1(a-m+1)-\psi _1(a+1)\big)+\frac{1}{2} \big(2 \psi _1(a+1) \big(\psi _0(a-m+1)-\psi _0(m+1)+\psi _0(1)\big)-2 \psi _1(m+1) \psi _0(a-m+1)+\psi _2(a-m+1)-\psi _2(a+1)\big)+\psi _0(a+1) \big(\psi _1(1)-\psi _1(a+1)\big)\qquad\qquad\qquad\quad
\end{dmath*}
\end{dgroup}
\begin{dgroup}
\hypertarget{16}{}\begin{dmath*}\label{16}
 \sum _{\smash{k=1}}^m \frac{ \psi _0(k)\psi _0(a+1-k)}{a+1-k}= \frac{1}{2} \sum _{k=1}^m \bigg(\frac{\psi _0^2(a+1-k)}{k}-\frac{\psi _1(k+a-m)}{k}\bigg)+\frac{1}{2(a-m)^2}\big(\psi _0(a-m+1)-\psi _0(a+1)+\psi _0(m+1)-\psi _0(1)\big)+\frac{1}{2(a-m)}\big(\psi _1(a+1)-\psi _1(a-m+1)\big)+\frac{1}{4} \big(2 \psi _0(m+1) \big(\psi _1(a+1)-\psi _0^2(a-m+1)\big)+2 \big(\psi _0(a+1)-\psi _0(a-m+1)\big) \big(\psi _1(a+1)-\psi _1(a-m+1)\big)-2 \psi _0(1) \psi _1(a-m+1)-\psi _2(a-m+1)+2 \psi _0(1) \psi _0^2(a+1)+\psi _2(a+1)\big)
\end{dmath*}
\end{dgroup}
\hypertarget{17}{}\begin{dgroup}\label{17}
\begin{dmath*}
\sum _{\smash{k=1}}^m \frac{\psi _0(k) \psi _0(a+1-k)}{k}=\frac{1}{2}\sum _{\smash{k=1}}^m \bigg( \frac{\psi _0^2(k+a-m)}{k} +\frac{\psi _0^2(k)}{k+a-m}- \frac{\psi _1(k+a-m)}{k}+ \frac{\psi _1(k)}{k+a-m}\bigg)+\left(\frac{1}{a-m}-\psi _0(a+1)\right) \sum _{\smash{k=1}}^m \frac{\psi _0(k+a-m)}{k}+\frac{1}{(a-m)^2}\big(\psi _0(a-m+1)-\psi _0(a+1)+\psi _0(m+1)-\psi _0(1)\big)-\frac{1}{2 (a-m)}\big(2 \psi _0(a+1) \big(\psi _0(a-m+1)+\psi _0(m+1)-\psi _0(1)\big)-\psi _0^2(a-m+1)+\psi _1(a-m+1)-\psi _0^2(a+1)-\psi _1(a+1)\big)+\frac{1}{6} \big(3 \psi _0^2(a+1) \big(\psi _0(m+1)-\psi _0(a-m+1)\big)-3 \psi _0(a+1) \big(2 \psi _0(1) \psi _0(a-m+1)-\psi _0^2(a-m+1)+\psi _1(a-m+1)-\psi _1(a+1)+\psi _0^2(1)+\psi _1(1)\big)-\psi _0^3(a-m+1)+3 \psi _0(1) \psi _0^2(a-m+1)+3 \psi _1(a+1) \psi _0(m+1)-3 \psi _0(1) \psi _1(a-m+1)+3 \psi _0(a-m+1) \big(\psi _1(a-m+1)-\psi _1(a+1)+\psi _0^2(m+1)+\psi _1(m+1)\big)-\psi _2(a-m+1)+\psi _0^3(a+1)+\psi _2(a+1)\big)
\end{dmath*}
\end{dgroup}
\hypertarget{18}{}\begin{dgroup}\label{18}
\begin{dmath*}
\sum _{\smash{k=1}}^m  \frac{\psi _0(k)\psi _0(a+1-k) }{m+1-k}=\frac{1}{2}\sum _{\smash{k=1}}^m \bigg( \frac{\psi _0^2(a+1-k)}{k}-  \frac{\psi _0^2(k)}{k+a-m}+  \frac{\psi _1(k+a-m)}{k}\\-  \frac{\psi _1(k)}{k+a-m}\bigg)-\left(\frac{1}{a-m}-\psi _0(a+1)\right)\\\times\sum _{\smash{k=1}}^m \frac{\psi _0(k+a-m)}{k}+\frac{1}{2 (a-m)^2}\big(3 \big(\psi _0(a+1)-\psi _0(a-m+1)-\psi _0(m+1)+\psi _0(1)\big)\big)-\frac{1}{2 (a-m)}\big(\psi _0^2(a+1)+2 \big(\psi _0(1)-2 \psi _0(m+1)\big) \psi _0(a+1) -\psi _0^2(a-m+1)+2 \psi _0(1) \psi _0(a-m+1)+2 \big(\psi _0(m+1)-\psi _0(1)\big) \psi _0(m+1)+2 \psi _1(m+1) -2 \psi _1(1)\big) +\frac{1}{12} \big(6 \psi _0(a-m+1)  \big(\big(-\psi _0(a+1)\\+\psi _0(m+1)-2 \psi _0(1)\big) \big(\psi _0(m+1)-\psi _0(a+1)\big)+\psi _1(m+1)-2 \psi _1(1)\big)-4 \psi _0^3(a+1)+6 \psi _0(1) \psi _0^2(a+1)+6 \psi _0^2(m+1) \psi _0(a+1)+6 \psi _1(a-m+1) \psi _0(a+1)-6 \psi _0(m+1) \big(\psi _0^2(a+1)+\psi _1(a-m+1)\big)+6 \psi _1(m+1) \psi _0(a+1)-6 \psi _0(a+1) \psi _1(a+1)-\psi _2(a+1)-2 \psi _0^3(a-m+1)+6 \psi _0(1) \psi _1(a-m+1)+\psi _2(a-m+1)\big)
\end{dmath*}
\end{dgroup}
\hypertarget{19}{}\begin{dgroup}\label{19}
\begin{dmath*}
\sum _{\smash{k=1}}^m \frac{\psi _0(k)\psi _0(k+a-m) }{m+1-k}=\frac{1}{2}\sum _{\smash{k=1}}^m \bigg( \frac{\psi _0^2(a+1-k)}{k}+ \frac{\psi _0^2(k+a-m)}{k}+ \frac{\psi _0^2(k)}{k+a-m}+ \frac{\psi _1(k)}{k+a-m}\bigg)+\frac{1}{a-m} \\\times\sum _{\smash{k=1}}^m \frac{\psi _0(k+a-m)}{k}+\frac{1}{2 (a-m)^2}\big(\psi _0(a-m+1)-\psi _0(a+1)+\psi _0(m+1)-\psi _0(1)\big)+\frac{1}{2 (a-m)}\big(\psi _0^2(a-m+1)-\psi _0^2(a+1)\big)+\frac{1}{12} \big(6 \psi _0^2(a+1) \big(\psi _0(a-m+1)+\psi _0(1)\big)+6 \psi _0(a+1) \\ \times \big(-2 \psi _0(m+1) \big(\psi _0(a-m+1)+\psi _0(1)\big)+\psi _1(a-m+1)-\psi _1(a+1)+\psi _0^2(m+1)+\psi _1(m+1)-2 \psi _1(1)\big)-2 \psi _0^3(a-m+1)+6 \psi _0(1) \psi _0^2(a-m+1)+6 \psi _0(m+1) \big(\psi _1(a+1)-\psi _1(a-m+1)\big)+6 \big(\psi _0^2(m+1)+\psi _1(m+1)\big) \psi _0(a-m+1)+\psi _2(a-m+1)-4 \psi _0^3(a+1)-\psi _2(a+1)\big)
\end{dmath*}
\end{dgroup}
\hypertarget{20}{}\begin{dgroup}\label{20}
\begin{dmath*}
\sum _{\smash{k=1}}^m \frac{\psi _0(k) \psi _0(k+b)}{k}=-\frac{1}{2}\sum _{\smash{k=1}}^m \bigg(\frac{\psi _1(k)}{k+b}+ \frac{\psi _0^2(k)}{k+b}\bigg)-\frac{1}{b}\sum _{\smash{k=1}}^m \frac{\psi _0(k+b)}{k}+\frac{1}{b^2}\big(\psi _0(b+m+1)-\psi _0(b+1)-\psi _0(m+1)+\psi _0(1)\big)+\frac{1}{2 b}\big(2 \psi _0(m+1) \psi _0(b+m+1)-2 \psi _0(1) \psi _0(b+1)-\psi _0^2(m+1)-\psi _1(m+1)+\psi _0^2(1)+\psi _1(1)\big)\\+\frac{1}{2} \big(\big(\psi _0^2(m+1)+\psi _1(m+1)\big) \psi _0(b+m+1)-\big(\psi _0^2(1)+\psi _1(1)\big) \psi _0(b+1)\big)
\end{dmath*}
\end{dgroup}
\hypertarget{21}{}\begin{dgroup}\label{21}
\begin{dmath*}
\sum _{\smash{k=1}}^m \frac{\psi _0(k) \psi _0(k+b)}{k+b}=-\frac{1}{2} \sum _{\smash{k=1}}^m \bigg(\frac{\psi _0^2(k+b)}{k}+ \frac{\psi _1(k+b)}{k}\bigg)-\frac{1}{b}\sum _{\smash{k=1}}^m \frac{\psi _0(k+b)}{k}+\frac{1}{2 b}\psi _0^2(b+m+1)+\psi _1(b+m+1)-\psi _0^2(b+1)-\psi _1(b+1)+\frac{1}{2} \big(\psi _0(m+1) \big(\psi _0^2(b+m+1)+\psi _1(b+m+1)\big)-\psi _0(1) \psi _0^2(b+1)-\psi _0(1) \psi _1(b+1)\big)
\end{dmath*}
\end{dgroup}
\makeatletter
\renewcommand{\theequation}{A.\number\numexpr\value{equation}-1\relax}
\makeatother
\begin{eqnarray}  
&&  \quad~ \bignumber \! \sum _{\smash{k=1}}^m\bigg( \frac{\psi_0 (k+b) \psi_0 (k+c)}{k+a}+ \frac{\psi_0 (k+a) \psi_0 (k+c)}{k+b}+ \frac{\psi_0 (k+a) \psi_0 (k+b)}{k+c}\bigg)
       \nonumber\label{formula}\\& =&\bigg(\frac{1}{b-a}+\psi_0 (a)\bigg) \sum _{k=1}^m \frac{\psi_0 (k+c)}{k+b}+\bigg(\frac{1}{c-a}+\psi_0 (a)\bigg) \sum _{k=1}^m \frac{\psi_0 (k+b)}{k+c}\nonumber\\
       &&+\bigg(\frac{1}{a-b}+\psi_0 (b)\bigg) \sum _{k=1}^m \frac{\psi_0 (k+c)}{k+a}+\bigg(\frac{1}{c-b}+\psi_0 (b)\bigg) \sum _{k=1}^m \frac{\psi_0 (k+a)}{k+c} \nonumber\\
       &&+\bigg(\!\frac{1}{a-c}+\psi_0 (c+m)\!\bigg)\!\sum _{k=1}^m \frac{\psi_0 (k+b)}{k+a}+\bigg(\!\frac{1}{b-c}+\psi_0 (c+m)\!\bigg) \!\sum _{k=1}^m \frac{\psi_0 (k+a)}{k+b} \nonumber\\
       &&+ \bigg(\frac{1}{c-b}-\frac{1}{c+m}\bigg) \psi_0 (a)\psi_0 (b+m)+\bigg(\frac{1}{c-a}-\frac{1}{c+m}\bigg) \psi_0 (a+m)\psi_0 (b) \nonumber\\
       &&+\psi_0 (a) \psi_0 (b) \psi_0 (c+m) -\psi_0 (a) \psi_0 (b+m) \psi_0 (c+m)+\psi_0 (a) \psi_0 (b) \psi_0 (c)\nonumber\\
       &&-\psi_0 (b) \psi_0 (a+m) \psi_0 (c+m)+\frac{(a+m)  (a-b-c-m)}{(a-b) (a-c) (b+m) (c+m)}\psi_0 (a+m)\nonumber\\
       &&  +\frac{1}{b-a}\psi_0 (a+m) \psi_0 (c+m)+\frac{(b+m)(a-b+c+m)}{(a-b) (a+m) (b-c) (c+m)} \psi_0 (b+m)\nonumber\\
       &&+\frac{1}{a-b}\psi_0 (b+m) \psi_0 (c+m)+\frac{(c+m)  (a+b-c+m)}{(a-c) (a+m) (c-b) (b+m)}\psi_0 (c+m) \nonumber\\
       &&+\frac{1}{c+m}\psi_0 (a+m) \psi_0 (b+m)+\bigg(\frac{1}{a-c}+\frac{1}{b-c}+\frac{1}{c}\bigg)\psi_0 (a) \psi_0 (b) \nonumber\\
       &&+\bigg(\frac{1}{b-a}+\frac{1}{a-c}-\frac{1}{a+m}+\frac{1}{a}\bigg)\psi_0 (b) \psi_0 (c+m)+\frac{1}{c-b}\psi_0 (a) \psi_0 (c) \nonumber\\
       &&+\bigg(\frac{1}{a-b}+\frac{1}{b-c}-\frac{1}{b+m}+\frac{1}{b}\bigg)\psi_0 (a) \psi_0 (c+m)+\frac{1}{c-a}\psi_0 (b) \psi_0 (c)\nonumber\\&&+\frac{a  (-a+b+c)}{b c (a-b) (a-c)}\psi_0 (a)-\frac{b  (a-b+c)}{a c (a-b) (b-c)}\psi_0 (b)+\frac{c  (a+b-c)}{a b (a-c) (b-c)}\psi_0 (c)\nonumber
\end{eqnarray}

\begin{eqnarray}
 && \qquad\bignumber \sum _{k=1}^m\left( \left(\frac{1}{k+a}+\frac{1}{k+b}\right)\left(\psi_0 (k+a+b+m)- \psi_0 (k+a+b)\right)\right)\nonumber\label{23}\qquad\quad~\\
 &=&\frac{m }{a(a+ m)}\left(\psi _0(b+m+1)- \psi _0(b+1)\right)+\frac{m }{b (b+m)}\big(\psi _0(a+m+1)\nonumber\\
 &&-\psi _0(a+1)\big)-\psi _0(b+1) \psi _0(a+m+1)-\psi _0(a+1) \psi _0(b+m+1)\nonumber\\
 &&+\psi _0(a+m+1) \psi _0(b+m+1)-\left(\frac{1}{a+m}+\frac{1}{a}+\frac{1}{b+m}+\frac{1}{b}\right)\nonumber\\
 &&\times  \psi _0(a+b+m+1)+\frac{a+b+2 m}{(a+m) (b+m)} \psi _0(a+b+2 m+1)\nonumber\\
 &&+\psi _0(a+1) \psi _0(b+1)+\left(\frac{1}{a}+\frac{1}{b}\right) \psi _0(a+b+1)\nonumber
\end{eqnarray}

\begin{eqnarray}
&&\qquad \!\bignumber\!\!\sum _{\smash{k=1}}^m\! \left(\frac{\psi _1(k+2 \alpha+m)- \psi _1(k+2 \alpha)}{k}-\frac{\psi _1(k+2 \alpha)}{k+2 \alpha+m}+ \frac{\psi _1(k+2 \alpha+m)}{k+2 \alpha}\!\right)\!\label{24}\nonumber\\
&=&\psi _0(1) \psi _1(2 \alpha +1)+\frac{1}{2 \alpha }\psi _1(2 \alpha +1)-\psi _0(2 \alpha +1) \psi _1(2 \alpha +1)+\frac{1}{2} \psi _2(2 \alpha +1)\nonumber\\
&&-\frac{4 \alpha ^2+m^2+6 \alpha  m}{2 \alpha  m^2+4 \alpha ^2 m} \psi _1(2 \alpha +m+1)
-\!\left(\!\frac{1}{m^2}+\frac{1}{(2 \alpha +m)^2}\!\right)\! \psi _0(2 \alpha +2 m+1)\nonumber\\
&&+\frac{1}{4 \alpha ^2 m^2 (2 \alpha +m)^2}\!\left(8 \alpha ^2 m\! \left(2 \alpha ^2+m^2+3 \alpha  m\right)\!\psi _1(2 \alpha +2 m+1) \right. \nonumber\\
&&-\left(4 \alpha ^2+m^2\right)(2 \alpha +m)^2\psi _0(2 \alpha +1)+\left(32 \alpha ^4+m^4+4 \alpha  m^3+16 \alpha ^2 m^2\right.\nonumber\\
&&\left.\left.+32 \alpha ^3 m\right)\psi _0(2 \alpha +m+1)\right)+ \!\left(\frac{1}{4 \alpha ^2}-\frac{1}{(2 \alpha +m)^2}\!\right)\big(\psi _0(1)-\psi _0(m+1)\big)\nonumber\\
&&-\psi _0(1) \psi _1(2 \alpha +m+1)-\psi _1(2 \alpha +1) \psi _0(m+1)-\frac{1}{2} \psi _2(2 \alpha +m+1)\nonumber\\
&&+\psi _0(m+1) \psi _1(2 \alpha +m+1)+\psi _0(2 \alpha +m+1) \psi _1(2 \alpha +m+1)\nonumber\\
&&-\psi _0(2 \alpha +2 m+1) \psi _1(2 \alpha +m+1)+\psi _1(2 \alpha +1) \psi _0(2 \alpha +m+1)\nonumber
\end{eqnarray}
\begin{eqnarray}
&&\qquad \bignumber \sum _{\smash{k=1}}^m\bigg( \frac{\psi _1(k)}{2 \alpha +k}- \frac{\psi _1(2 \alpha +k)}{2 \alpha +k}- \frac{\psi _1(k)}{2 \alpha +k+m}+ \frac{\psi _1(2 \alpha +k)}{2 \alpha +k+m}\bigg)\nonumber\label{25}\\
&=&-\psi _0(1) \psi _1(2 \alpha +1)-\left(\psi _1(m+1) \left(\psi _0(2 \alpha +1)-2 \psi _0(2 \alpha +m+1)\right)\right)\nonumber\\
&&-\psi _1(m+1) \psi _0(2 \alpha +2 m+1)+\psi _0(1) \psi _1(2 \alpha +m+1)\nonumber\\
&&-\psi _0(m+1) \psi _1(2 \alpha +m+1)-\psi _0(2 \alpha +m+1) \psi _1(2 \alpha +m+1)\nonumber\\
&&+\psi _0(2 \alpha +2 m+1) \psi _1(2 \alpha +m+1)+\psi_1(2\alpha +1) \psi _0(m+1)\nonumber\\
&&+\psi _1(2 \alpha +1) \left(\psi _0(2 \alpha +1)-\psi_0(2\alpha +m+1)\right)\nonumber\qquad\qquad\qquad\qquad\qquad\qquad\quad
\end{eqnarray}

\section{Correlation Functions}\label{HAHD}
In this appendix, the explicit expression of the three-point density function utilized in formulating cumulant integrals in Section~\ref{sec:3.1}
is provided:
\begin{equation*}
h_{3}(x,y,z)=\frac{1}{8m(m-1)(m-2)}\left(h_{\text{A}}+h_{\text{B}}+h_{\text{C}}+h_{\text{D}}\right)
\end{equation*}
that can be found, for example, in~\cite{LW21}\counterwithin{equation}{section}, where
\begin{eqnarray}
h_{\text{A}} &=& (K_{01}(x,x)+K_{10}(x,x))(K_{01}(y,y)+K_{10}(y,y))\nonumber\\&&\times(K_{01}(z,z)+K_{10}(z,z))\label{eq:hA}\\
h_{\text{B}} &=& -2(K_{01}(x,x)+K_{10}(x,x))(K_{01}(y,z)K_{01}(z,y)+K_{10}(y,z)K_{10}(z,y) \nonumber \\
&&+K_{00}(y,z)K_{11}(y,z)+K_{00}(z,y)K_{11}(z,y))-2(K_{01}(y,y)+K_{10}(y,y))\nonumber \\
&&\times(K_{01}(x,z)K_{01}(z,x)+K_{10}(x,z)K_{10}(z,x)+K_{00}(x,z)K_{11}(x,z)\nonumber \\
&&+K_{00}(z,x)K_{11}(z,x))-2(K_{01}(z,z)+K_{10}(z,z))(K_{01}(x,y)K_{01}(y,x)\nonumber \\
&&+K_{10}(x,y)K_{10}(y,x)+K_{00}(x,y)K_{11}(x,y)+K_{00}(y,x)K_{11}(y,x))\label{eq:hB} \\
h_{\text{C}} &=& 2(K_{00}(x,y)K_{01}(y,z)K_{11}(x,z)+K_{00}(x,y)K_{10}(z,x)K_{11}(z,y) \nonumber \\
&&+K_{00}(y,x)K_{01}(x,z)K_{11}(y,z)+K_{00}(y,x)K_{10}(z,y)K_{11}(z,x) \nonumber \\
&&+K_{00}(x,z)K_{01}(z,y)K_{11}(x,y)+K_{00}(x,z)K_{10}(y,x)K_{11}(y,z)\nonumber \\
&&+K_{00}(z,x)K_{01}(x,y)K_{11}(z,y)+K_{00}(z,x)K_{10}(y,z)K_{11}(y,x)\nonumber \\
&&+K_{00}(y,z)K_{01}(y,x)K_{11}(x,z)+K_{00}(y,z)K_{10}(x,z)K_{11}(y,x)\nonumber \\
&&+K_{00}(z,y)K_{01}(z,x)K_{11}(x,y)+K_{00}(z,y)K_{10}(x,y)K_{11}(z,x)\nonumber \\
&&-K_{00}(x,y)K_{01}(x,z)K_{11}(y,z)-K_{00}(x,y)K_{10}(z,y)K_{11}(z,x)\nonumber \\
&&-K_{00}(y,x)K_{01}(y,z)K_{11}(x,z)-K_{00}(y,x)K_{10}(z,x)K_{11}(z,y)\nonumber \\
&&-K_{00}(x,z)K_{01}(x,y)K_{11}(z,y)-K_{00}(x,z)K_{10}(y,z)K_{11}(y,x)\nonumber \\
&&-K_{00}(z,x)K_{01}(z,y)K_{11}(x,y)-K_{00}(z,x)K_{10}(y,x)K_{11}(y,z)\nonumber \\
&&-K_{00}(y,z)K_{01}(z,x)K_{11}(x,y)-K_{00}(y,z)K_{10}(x,y)K_{11}(z,x) \nonumber \\
&&-K_{00}(z,y)K_{01}(y,x)K_{11}(x,z)-K_{00}(z,y)K_{10}(x,z)K_{11}(y,x))\label{eq:hC} \\
h_{\text{D}} &=& 2(K_{01}(x,y)K_{01}(y,z)K_{01}(z,x)+K_{01}(x,z)K_{01}(z,y)K_{10}(x,y) \nonumber \\
&&+K_{01}(x,z)K_{01}(y,x)K_{10}(y,z)+K_{01}(y,x)K_{01}(z,y)K_{10}(z,x)\nonumber \\
&&+K_{01}(x,y)K_{10}(x,z)K_{10}(z,y)+K_{01}(y,z)K_{10}(x,z)K_{10}(y,x) \nonumber \\
&&+K_{01}(z,x)K_{10}(y,x)K_{10}(z,y)+K_{10}(x,y)K_{10}(y,z)K_{10}(z,x)).\label{eq:hD}
\end{eqnarray}

\section{Re-summation Functions for Anomalies}\label{secA2}
\counterwithin{equation}{section}
We list here re-summation functions $G(m)$ for the anomalies with negative summation variables used in the derivation of summation identities~(\hyperlink{3}{\ref*{3}}),~(\hyperlink{4}{\ref*{4}}), (\hyperlink{10}{\ref*{10}})--(\hyperlink{16}{\ref*{16}}),~(\hyperlink{18}{\ref*{18}}), and~(\hyperlink{19}{\ref*{19}}):
\begin{eqnarray}
&&\sum\limits_{\smash{k=1}}^m \frac{\psi _0(b+m+1-k)}{k} \text{ for~(\hyperlink{3}{\ref*{3}})} \\ 
&&\sum\limits_{\smash{k=1}}^m \frac{\psi _0(m+1-k)}{k+b} \text{ for~(\hyperlink{4}{\ref*{4}})} \\
&&\sum\limits_{\smash{k=1}}^m \frac{\psi _0(m+1-k)}{(k+b)^2} \text{ for~(\hyperlink{10}{\ref*{10}})} \\
&&\sum\limits_{\smash{k=1}}^m \frac{\psi _0^2(m+1-k)}{k+b} \text{ for~(\hyperlink{11}{\ref*{11}})} \\
&&\sum\limits_{\smash{k=1}}^m \frac{\psi _1(b+m+1-k)}{k} \text{ for~(\hyperlink{12}{\ref*{12}})} \\
&&\sum\limits_{\smash{k=1}}^m \frac{\psi _1(m+1-k)}{k+b} \text{ for~(\hyperlink{14}{\ref*{14}})} \\
&&\sum\limits_{\smash{k=1}}^m \frac{\psi _0(b+m+1-k)}{k^2} \text{ for~(\hyperlink{15}{\ref*{15}})} \\
&&\sum\limits_{\smash{k=1}}^m \frac{\psi _0(m+1-k) \psi _0(k+b)}{k+b} \text{ for~(\hyperlink{16}{\ref*{16}})} \\ 
&&\sum\limits_{\smash{k=1}}^m \frac{\psi _0(m+1-k) \psi _0(k+b)}{k} \text{ for~(\hyperlink{18}{\ref*{18}})} \label{eq:f18} \\
&&\sum\limits_{\smash{k=1}}^m \frac{\psi _0(k) \psi _0(k+b)}{m+1-k} \text{ for~(\hyperlink{19}{\ref*{19}})}. \label{eq:f19}
\end{eqnarray}   

\begin{acknowledgements}
The work of Lu Wei was supported by the U.S. National Science Foundation (2306968) and the U.S. Department of Energy (DE-SC0024631).
\end{acknowledgements}

\providecommand{\noopsort}[1]{}\providecommand{\singleletter}[1]{#1}%

\bibliographystyle{spmpsci}      

\end{document}